\documentclass[journal=jctc,manuscript=article]{achemso}

\usepackage{amssymb}
\usepackage{amsthm}
\usepackage{amsmath}
\usepackage{braket}
\usepackage{graphicx}
\usepackage{hyperref}
\usepackage{multirow}
\usepackage{xcolor}

\setkeys{acs}{maxauthors=0}

\SectionNumbersOn

\newcommand{\s}{{\textcolor{white}{1}}}

\author{Victor Wen-zhe Yu}
\author{Marco Govoni}
\email{mgovoni@anl.gov}
\altaffiliation{Pritzker School of Molecular Engineering, The University of Chicago, Chicago, Illinois 60637, USA}
\affiliation{Materials Science Division, Argonne National Laboratory, Lemont, Illinois 60439, USA}

\title{GPU Acceleration of Large-Scale Full-Frequency GW Calculations}


\begin{document}

\begin{abstract}
Many-body perturbation theory is a powerful method to simulate electronic excitations in molecules and materials starting from the output of density functional theory calculations. By implementing the theory efficiently so as to run at scale on the latest leadership high-performance computing systems it is possible to extend the scope of GW calculations. We present a GPU acceleration study of the full-frequency GW method as implemented in the WEST code. Excellent performance is achieved through the use of (i) optimized GPU libraries, e.g., cuFFT and cuBLAS, (ii) a hierarchical parallelization strategy that minimizes CPU-CPU, CPU-GPU, and GPU-GPU data transfer operations, (iii) nonblocking MPI communications that overlap with GPU computations, and (iv) mixed precision in selected portions of the code. A series of performance benchmarks has been carried out on leadership high-performance computing systems, showing a substantial speedup of the GPU-accelerated version of WEST with respect to its CPU version. Good strong and weak scaling is demonstrated using up to 25920 GPUs. Finally, we showcase the capability of the GPU version of WEST for large-scale, full-frequency GW calculations of realistic systems, e.g., a nanostructure, an interface, and a defect, comprising up to 10368 valence electrons.
\end{abstract}

\section{Introduction}
\label{sec:introduction}

First-principles simulations of materials have become mainstream computational instruments to understand energy conversion processes in several areas of materials science and chemistry, including, for instance, applications to photovoltaics and photocatalysis. Simulations using the Kohn-Sham density functional theory (KS-DFT)~\cite{dft_hohenberg_1964,dft_kohn_1965} are widely adopted to computationally predict the structures and properties of molecules and materials in their ground state. However, KS-DFT methods fail to provide an accurate description of electrons in excited states. The GW method, formulated within the context of many-body perturbation theory~\cite{gw_strinati_1988,gw_golze_2019}, has been established as the main method to improve the electronic structure obtained with DFT and describe excited states. The GW self-energy was initially proposed by Hedin~\cite{hedin_hedin_1965} as a numerically manageable approximation to the complex many-body nature of electron-electron interactions. The earliest applications of the GW method to electronic structure of semiconductors and insulators obtained with DFT date back to the 1980s~\cite{gw_strinati_1980,gw_strinati_1982,gw_hybertsen_1985,gw_hybertsen_1986,gw_aryasetiawan_1998}. Conventional GW implementations, currently available in several electronic structure codes, have a computational cost that scales as $\mathcal{O}(N^4)$ with respect to the system size $N$, limiting the tractable size of GW calculations. Method development and code optimization have been active areas of research in order to push the scope of applicability of such GW implementations to large systems. Formulations with cubic scaling algorithms~\cite{cubic_foerster_2011,cubic_liu_2016,cubic_wilhelm_2018,adf_forster_2020,low_wilhelm_2021,cubic_duchemin_2021,low_forster_2021} or stochastic methods~\cite{stochastic_neuhauser_2014,stochastic_vlcek_2017,stochastic_vlcek_2018,stochastic_brooks_2020,stochastic_romanova_2022} have been proposed, albeit at the cost of introducing expensive numerical integration operations or stochastic errors, respectively.

The rise of heterogeneous computing has substantially increased the throughput available in leadership high-performance computing (HPC) systems to hundreds of PFLOP/s (peta floating-point operations per second), and we are currently witnessing the transition to the exascale. On the current release (November 2021) of the TOP500 list~\cite{top500}, seven of the top ten supercomputers have graphics processing units (GPUs), including Summit, the world's second fastest computer powered by 27648 NVIDIA V100 GPUs. GPU devices consist of hundreds to thousands of cores that operate at a relatively low frequency and can perform parallel computational tasks in a more energy efficient way than by central processing units (CPUs). This sets tremendous opportunities for first-principles simulations, including the ability to carry out GW calculations at unprecedented scales. However, most software packages in the electronic structure community were initially written to target traditional CPUs, with parallelization primarily managed by the message passing interface (MPI). The migration to accelerated, heterogeneous computing typically requires a redesign of the code to fully harness the parallelism of modern GPUs. GPU acceleration has been reported by a number of electronic structure theory and quantum chemistry software packages~\cite{abinit_gonze_2016,bigdft_ratcliff_2018,cp2k_kuhne_2020,fhiaims_huhn_2020,nwchem_apra_2020,octopus_tancognedejean_2020,qe_giannozzi_2020,terachem_seritan_2020}. For the GW method in particular, the Gaussian-orbital-based VOTCA-XTP code~\cite{votcaxtp_tirimbo_2020} and the plane-wave-based Yambo~\cite{yambo_sangalli_2019} code can perform GPU-accelerated GW calculations of molecules and materials. The plane-wave-based BerkeleyGW code was recently ported to run on GPUs to carry out a large-scale GW calculation for a silicon model consisting of 10968 valence electrons using a generalized plasmon-pole model to approximate retardation effects~\cite{berkeleygw_delben_2020}.

In this paper, we present the GPU porting of the WEST code~\cite{west_govoni_2015,west_website}, a plane-wave pseudopotential implementation of the full-frequency G$_0$W$_0$ method. In addition to featuring a massive parallelization, demonstrated using over $\sim$500000 CPU cores in reference~\citenum{west_govoni_2015}, WEST uses techniques to help prevent computational and memory bottlenecks for large systems; for instance, it represents the density-density response functions in a compact basis set, eliminating the need to store and manipulate large matrices. The slowly converging sum over empty KS states, commonly encountered in most GW codes, is avoided completely in WEST. WEST carries out a full integration over the frequency domain, removing the need of approximating retardation effects with plasmon-pole models. The accuracy of the full-frequency implementation in WEST was recently assessed, verifying the implementation by comparing the results obtained with WEST with the results of all-electron codes~\cite{gw100_govoni_2018}. The WEST code has been used to study excited states for a variety of systems, including molecules, nanoparticles, two-dimensional (2D) materials, solids, defects in solids, liquids, amorphous, and solid/liquid interfaces~\cite{gw100_govoni_2018,westapp_seo_2016,westapp_gaiduk_2016,westapp_scherpelz_2016,westapp_pham_2017,westapp_seo_2017,westapp_gaiduk_2018,westapp_smart_2018,westapp_gerosa_2018,westapp_zheng_2019,westapp_yang_2019,westapp_kundu_2021,westapp_jin_2021}. Recent developments within WEST include the computation of electron-phonon self-energies~\cite{phonon_mcavoy_2018,phonon_yang_2021} and absorption spectra~\cite{absorption_nguyen_2019,absorption_dong_2021} and the formulation of a quantum embedding approach~\cite{embedding_ma_2020a,embedding_ma_2020b,embedding_ma_2021}. The GPU porting of WEST aims to further advance the simulation of electronic excitations in large, complex materials on a variety of GPU-powered, pre-exascale and exascale HPC systems. The strategy reported here is general and can be applied to other GW codes.

The rest of the paper is organized as follows. In section~\ref{sec:g0w0}, we briefly review the G$_0$W$_0$ theory and the current state of the art. In section~\ref{sec:west}, we summarize the implementation in the WEST code. We then introduce the GPU porting of WEST in section~\ref{sec:gpu}, elaborating on several optimization strategies that help maximize the efficiency of the code, especially when running on a large number of GPUs. The performance of the newly developed GPU version of WEST is discussed in section~\ref{sec:performance} with a series of benchmarks, demonstrating excellent performance and scalability on leadership HPC systems. In section~\ref{sec:large}, we report three examples of large full-frequency G$_0$W$_0$ calculations. Our conclusions are given in section~\ref{sec:conclusions}.

\section{G$_0$W$_0$ Method}
\label{sec:g0w0}

\subsection{Theory}

In KS-DFT~\cite{dft_hohenberg_1964,dft_kohn_1965}, the ground state of a system of interacting electrons in the external field of the ions may be obtained by solving the KS set of single-particle equations
\begin{equation}
\label{eq:ks}
h_\text{KS}^\sigma \psi_{i\sigma} = \varepsilon_{i\sigma} \psi_{i\sigma} \,,
\end{equation}
where $\psi_{i\sigma}$ and $\varepsilon_{i\sigma}$ correspond to the wave function and energy of the $i$th KS state in the $\sigma$ spin channel, respectively. The KS Hamiltonian, $h_\text{KS}^\sigma$, includes the single-particle kinetic energy operator, $t_\text{s}$, and the Hartree, external (ionic), and exchange-correlation potential operators $v_\text{H}$, $v_\text{ext}$, and $v_\text{xc}^\sigma$, respectively. Throughout the paper we focus on large systems that do not require $k$-point sampling, therefore we omit $k$-point indices for simplicity.

Quasiparticle (QP) states may be obtained by solving the following Dyson-like equation
\begin{equation}
\label{eq:qp}
h_\text{QP}^\sigma \psi_{i\sigma}^\text{QP} = \varepsilon_{i\sigma}^\text{QP} \psi_{i\sigma}^\text{QP} \,,
\end{equation}
where the QP Hamiltonian, $h_\text{QP}^\sigma$, is obtained from the KS Hamiltonian by replacing the exchange and correlation potential with the electron self-energy $\Sigma$. The latter is a frequency-dependent and nonlocal operator that may be expressed in a compact form as:
\begin{equation}
\label{eq:self_energy}
\Sigma = i G W \Gamma \,,
\end{equation}
where $G$, $W$, and $\Gamma$ are the Green's function, the screened Coulomb interaction, and the vertex operator, respectively. $\Sigma$ may be computed by solving Hedin's equations self-consistently~\cite{hedin_hedin_1965}. Within the G$_0$W$_0$ approximation~\cite{gw_hybertsen_1985,gw_hybertsen_1986,gw_aryasetiawan_1998}, $\Gamma$ is treated as the identity and the self-energy is evaluated not self-consistently as:
\begin{equation}
\label{eq:self_energy_gw}
\Sigma^\sigma (\boldsymbol{r}, \boldsymbol{r}^\prime; \omega) = i \int_{- \infty}^{+ \infty} \frac{\text{d} \omega^\prime}{2 \pi} G_0^\sigma (\boldsymbol{r}, \boldsymbol{r}^\prime; \omega + \omega^\prime)W_0 (\boldsymbol{r}, \boldsymbol{r}^\prime; \omega^\prime) \,.
\end{equation}
The KS states and energies may be used to evaluate all terms in the right-hand side (RHS) of equation~\ref{eq:self_energy_gw}, i.e., the non-self-consistent Green's function, $G^\sigma_0(\omega)=(\omega-h^\sigma_\text{KS})^{-1}$, and the screened Coulomb potential, $W_0=v+v^{1/2}\bar{\chi} v^{1/2}$, where $\bar{\chi}$ is the symmetrized density-density response function of the system. To obtain the latter, the irreducible density-density response function, $\chi_0=iG_0G_0$, is first evaluated; second, $\bar{\chi}$ is obtained within the random phase approximation (RPA) using a Dyson recursive equation, $\bar{\chi} = \bar{\chi}_0 + \bar{\chi}_0 \bar{\chi}$, where $\bar{\chi}_0=v^{1/2}\chi_0 v^{1/2}$.

Once the self-energy is obtained, QP energies are found using perturbation theory starting from the solution of equation~\ref{eq:ks}:
\begin{equation}
\label{eq:e_qp}
\begin{split}
\varepsilon_{i\sigma}^\text{QP} & = \varepsilon_{i\sigma} + \braket{\psi_{i\sigma} | h_\text{QP}^\sigma - h_\text{KS}^\sigma | \psi_{i\sigma}} \\
& = \varepsilon_{i\sigma} + \braket{\psi_{i\sigma} | \Sigma^\sigma (\varepsilon_{i\sigma}^\text{QP}) - v_\text{xc}^\sigma | \psi_{i\sigma}} \,.
\end{split}
\end{equation}

The frequency integration in equation~\ref{eq:self_energy_gw} can be evaluated numerically using the contour deformation technique~\cite{contour_godby_1988,contour_lebegue_2003,abinit_gonze_2009}, i.e., by carrying out the integration in the complex plane along a contour that excludes the poles of $W_0$:
\begin{equation}
\braket{\psi_{i\sigma} | \Sigma^\sigma (\omega) | \psi_{i\sigma}} = \braket{\psi_{i\sigma} | \Sigma^\sigma_X | \psi_{i\sigma}} + I_{i\sigma}(\omega)+R_{i\sigma}(\omega) \,.
\label{eq:sigmacIR}
\end{equation}
The exchange self-energy, $\Sigma_X$, is obtained by replacing $W_0$ in equation~\ref{eq:self_energy_gw} with the frequency-independent bare Coulomb potential $v$. $I_{i\sigma}$ contains an integration along the imaginary axis, where $G_0$ and $W_0$ are both smooth functions
\begin{equation}
I_{i\sigma}(\omega) = - \int_{-\infty}^{+\infty} \frac{d\omega^\prime}{2\pi} \int d\boldsymbol{r}\int d\boldsymbol{r^\prime} \psi_{i\sigma}^\ast(\boldsymbol{r}) G_0^\sigma (\boldsymbol{r},\boldsymbol{r^\prime}; \omega + i\omega^\prime) W_p(\boldsymbol{r},\boldsymbol{r^\prime}; i\omega^\prime) \psi_{i\sigma}(\boldsymbol{r^\prime}) \,.
\label{eq:Ifac}
\end{equation}
The $R_{i\sigma}$ term contains the residues associated with the poles of the Green's function that may fall inside the chosen contour
\begin{equation}
R_{i\sigma}(\omega) = \sum_j f^{i\sigma}_{j\sigma} \int d\boldsymbol{r}\int d\boldsymbol{r^\prime} \psi_{i\sigma}^\ast(\boldsymbol{r})\psi_{j\sigma}(\boldsymbol{r}) W_p(\boldsymbol{r},\boldsymbol{r^\prime}; \varepsilon_{j\sigma}-\omega) \psi_{j\sigma}^\ast(\boldsymbol{r^\prime})\psi_{i\sigma}(\boldsymbol{r^\prime}) \,.
\label{eq:Rfac}
\end{equation}
We labeled $W_p$ the part of the screened Coulomb potential that depends on the frequency, i.e., $W_p = W_0 - v$, and we defined $f^{i\sigma}_{j\sigma} = \theta(\varepsilon_{j\sigma}-\varepsilon_F) \theta(\varepsilon^{QP}_{j\sigma}-\varepsilon_{j\sigma}) - \theta(\varepsilon_F-\varepsilon_{j\sigma}) \theta(\varepsilon_{j\sigma}-\varepsilon^{QP}_{j\sigma})$, where $\theta(x)$ is the Heaviside step function and $\varepsilon_F$ is the Fermi energy; a formal derivation may be found in reference~\citenum{west_govoni_2015}.

Quasiparticle energies obtained by solving equation~\ref{eq:e_qp} are used to compute charged excitations and yield an electronic structure that can be compared to direct and inverse photoelectron spectroscopies (UPS, XPS, ARPES)~\cite{gw_golze_2019,gw_onida_2002,gw_ping_2013}.

\subsection{State of the Art}

First-principles calculations using the G$_0$W$_0$ method are typically carried out after DFT and are computationally more demanding than the latter: the computational complexity of the GW method scales as $\mathcal{O}(N^4)$ with respect to the system size $N$, whereas DFT scales as $\mathcal{O}(N^3)$. In addition, several computational bottlenecks hinder the applicability of the G$_0$W$_0$ method to large systems containing thousands of valence electrons ($N_\text{occ}$). In the following, we focus the discussion on implementations of $G_0W_0$ with three-dimensional (3D) periodic boundary conditions using the plane-wave basis set where $N_\rho$ and $N_\psi$ are the number of plane-waves associated with the chosen kinetic energy cutoff for the density and the wave function, respectively. In the case where ionic potentials are described using norm-conserving pseudopotentials, we have $N_\rho \simeq 8 N_\psi$. A first computational bottleneck occurs when one wants to evaluate $G_0$ using its Lehmann representation or $\chi_0$ using the Adler-Wiser formula~\cite{dielectric_adler_1962,dielectric_wiser_1963}, i.e., in terms of the eigenvectors and eigenvalues of $h_\text{KS}^\sigma$. In this case, a summation over occupied states and empty states must be taken explicitly. The bottleneck is caused by the difficulty to fully diagonalize the KS Hamiltonian in its empty manyfold because $N_\psi \gg N_\text{occ}$. A second computational bottleneck occurs when one wants to evaluate $W_p$ at multiple frequencies, and the density-density response function is represented at each frequency by a large matrix with $N_\rho$ elements per axis. The systems discussed in this manuscript have millions of plane-waves, requiring the storage and manipulation of large matrices.

To make the simulations tractable, conventional implementations of the G$_0$W$_0$ method introduce additional parameters, e.g., $N_\text{empty} \ll N_{\psi}$ and $N_\chi \ll N_\rho$, to limit the number of empty states and the size of density-density response functions, respectively. However, these parameters, not present in the DFT calculation, show a slow convergence with respect to the size of the system. In addition, several implementations of the G$_0$W$_0$ method solve equation~\ref{eq:e_qp} with linearization using the on-the-mass shell approximation (i.e., $\Sigma$ is evaluated at the KS energy), or approximate the frequency-dependent dielectric screening using generalized plasmon-pole models~\cite{gw_hybertsen_1986,gw_aryasetiawan_1998,gpp_linden_1988,gpp_engel_1993,gpp_shaltaf_2008,gpp_stankovski_2011}. Such models are derived for homogeneous systems and commonly applied to heterogeneous systems without formal justification~\cite{gpp_larson_2013}. Reproducibility studies have shown that these approximations can be the source of discrepancies between different implementations~\cite{gw100_govoni_2018,gw100_setten_2015,gw100_maggio_2017,reproduce_rangel_2020}.

Method development aimed at improving the efficiency of full-frequency G$_0$W$_0$ calculations is the focus of current research. A few techniques have been developed in order to reduce the cost of the sum over empty states, including the extrapolar approximation~\cite{extrapolar_bruneval_2008,extrapolar_bruneval_2016}, the static remainder approach~\cite{static_kang_2010}, the effective energy technique~\cite{effective_berger_2010,effective_berger_2012}, the multipole approach~\cite{multipole_leon_2021}, and methods~\cite{west_govoni_2015,gw_reining_1997,dielectric_wilson_2008,dielectric_wilson_2009,gw_umari_2009,lanczos_umari_2010,gw_giustino_2010,gw_lambert_2013,pdep_nguyen_2012,pdep_pham_2013,gw_janssen_2015} based on density functional perturbation theory (DFPT)~\cite{dfpt_baroni_1987,dfpt_baroni_2001}. The stochastic formulation of G$_0$W$_0$~\cite{stochastic_neuhauser_2014,stochastic_vlcek_2017,stochastic_vlcek_2018}, which employs the time evolution of the occupied states, leads to an implementation that does not involve empty states, and its results are comparable to those obtained with the deterministic full frequency G$_0$W$_0$ method.

Implementations of the G$_0$W$_0$ method using plane-waves basis sets are available in the following codes: ABINIT~\cite{abinit_gonze_2016}, BerkeleyGW~\cite{berkeleygw_deslippe_2012}, GPAW~\cite{gpaw_huser_2013}, OpenAtom~\cite{openatom_kim_2019}, Quantum ESPRESSO~\cite{qe_giannozzi_2020}, SternheimerGW~\cite{sternheimergw_schlipf_2020}, VASP~\cite{vasp_kresse_1996}, WEST~\cite{west_govoni_2015} (this work), and Yambo~\cite{yambo_sangalli_2019}. Other implementations use Gaussian basis sets, such as Fiesta~\cite{fiesta_blase_2011}, MOLGW~\cite{molgw_bruneval_2016}, TURBOMOLE~\cite{turbomole_balasubramani_2020}, and VOTCA-XTP~\cite{votcaxtp_tirimbo_2020}, Slater type orbitals, such as ADF~\cite{adf_forster_2020}, numerical atomic orbitals, such as FHI-aims~\cite{fhiaims_ren_2021}, mixed Gaussian and plane-waves, such as CP2K~\cite{cp2k_kuhne_2020}, linearized augmented-plane-waves with local orbitals, such as Elk~\cite{elk}, Exciting~\cite{exciting_gulans_2014}, and FHI-gap~\cite{fhigap_jiang_2013}, and real-space grids, such as NanoGW~\cite{nanogw_tiago_2006} and StochasticGW~\cite{stochastic_neuhauser_2014}.

In the next section, we summarize the implementation of the full-frequency G$_0$W$_0$ method in the WEST code as presented in reference~\citenum{west_govoni_2015}, and we discuss the implications of design choices for calculations of large-scale system. In section~\ref{sec:gpu}, we present the porting of WEST to GPUs.

\section{G$_0$W$_0$ Implementation in the WEST Code}
\label{sec:west}

The open-source software package WEST (Without Empty STates)~\cite{west_govoni_2015,west_website} implements the full-frequency G$_0$W$_0$ method for large systems using 3D periodic boundary conditions. The underlying DFT electronic structure is obtained using the plane-wave pseudopotential method. Key features of the WEST code include (i) the use of algorithms to circumvent explicit summations over empty states, (ii) the use of a low-rank decomposition of response functions to avoid storage and inversion operations on large matrices, and (iii) the use of the Lanczos technique to facilitate the calculation of the density-density response at multiple frequencies.

\begin{figure}
\includegraphics[width=0.99\textwidth]{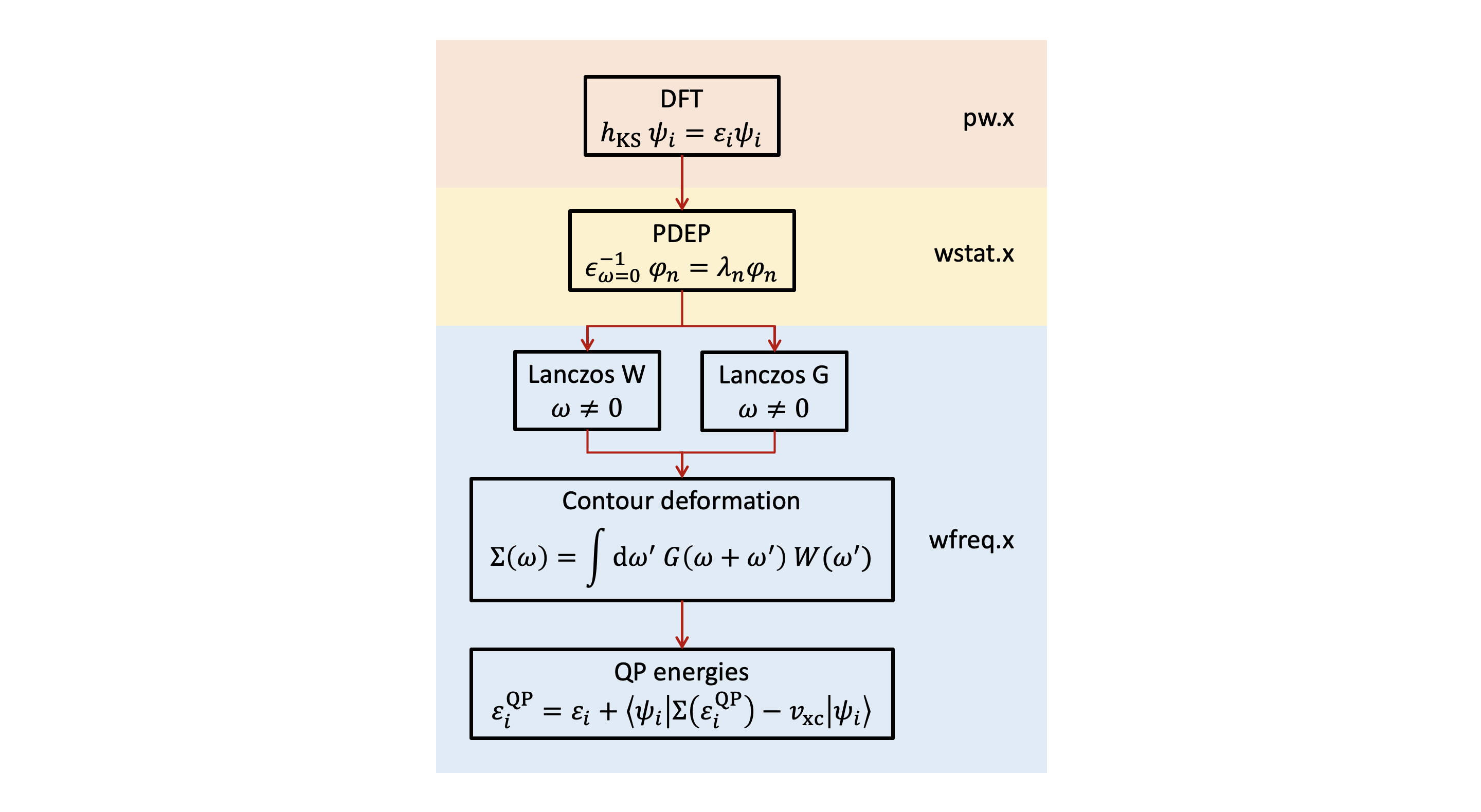}
\caption{Key steps to compute QP energies within the G$_0$W$_0$ approximation using the WEST code. The \texttt{pwscf} code (pw.x) in Quantum ESPRESSO is employed to compute the KS wave functions and energies at the DFT level. These quantities are input to the \texttt{wstat} code (wstat.x) in WEST, which generates the PDEP basis set by iteratively diagonalizing the static dielectric matrix at zero frequency. The \texttt{wfreq} code (wfreq.x) in WEST then uses the PDEP basis set to compute $G_0$ and $W_0$ at finite frequencies with the Lanczos algorithm. Frequency integration of the self-energy in equation~\ref{eq:self_energy_gw} is carried out with the contour deformation technique. Finally, the QP energies are solved using equation~\ref{eq:e_qp}.}
\label{fig:workflow}
\end{figure}

The complete workflow for computing QP energies with the WEST software is shown in figure~\ref{fig:workflow}, where the \texttt{pwscf} code (pw.x) in the Quantum ESPRESSO software suite~\cite{qe_giannozzi_2020,qe_giannozzi_2009,qe_giannozzi_2017} is used to carry out the ground state DFT calculation, the \texttt{wstat} code (wstat.x) in WEST constructs the projective dielectric eigenpotentials (PDEP) basis set that is used to obtain a low-rank representation of the density-density response function, and the \texttt{wfreq} code (wfreq.x) in WEST computes the QP energies. In the following, we describe each part of the workflow.

The ground-state electronic structure is obtained with DFT using semilocal or hybrid functionals. Although in this work we focus on spin-unpolarized and spin-polarized large systems of which the Brillouin zone can be sampled using the $\Gamma$ point, the WEST software supports simulations with $k$-points sampling and with noncollinear spin~\cite{westapp_scherpelz_2016}. Currently, only norm-conserving pseudopotentials are supported.

Starting from the output of DFT, the PDEP algorithm is used to find the leading eigenvectors of the symmetrized irreducible density-density response function $\bar{\chi}^0$ at zero frequency~\cite{dielectric_wilson_2008,dielectric_wilson_2009}. The eigenvectors of $\bar{\chi}^0(\omega=0)$, referred to as the PDEP basis set, are then used to construct a low-rank decomposition of the symmetrized reducible density-density response function $\bar{\chi}$ at finite frequencies. Finally, by using the PDEP basis set one may express $W_p$ in a separable form:
\begin{equation}
\label{eq:Wp}
W_p(\boldsymbol{r},\boldsymbol{r^\prime};\omega) = \Xi(\omega) + \frac{1}{\Omega} \sum_{nm}^{N_\text{PDEP}} \Lambda_{nm}(\omega) \tilde{\varphi}_n(\boldsymbol{r}) \tilde{\varphi}_m(\boldsymbol{r^\prime}) \,,
\end{equation}
where $\Lambda_{nm}$ are the matrix elements of the operator on the PDEP basis set, $\Xi$ takes into account the frequency-dependent long-range dielectric response, $\Omega$ is the volume of the simulation cell, and $\tilde{\varphi}_n$ are symmetrized eigenpotentials, i.e., $\tilde{\varphi}_m = v^{1/2}\varphi_m$. A formal derivation may be found in reference~\citenum{west_govoni_2015}.

The PDEP algorithm uses the Davidson method~\cite{davidson_davidson_1975} to find the leading eigenvectors of $\bar{\chi}^0$. This is done by first constructing an orthonormal set of $N_\text{PDEP}$ trial vectors $\{\varphi_j: j = 1,...,N_\text{PDEP}\}$. We then repeatedly apply $\bar{\chi}^0$ to the vectors of the set and expand the set by including the residues, until the set contains $N_\text{PDEP}$ leading eigenvectors of the operator. At each iteration of the Davidson algorithm, the result of the application of $\bar{\chi}^0$ on each vector of the set is obtained by computing the symmetrized density-density response of the system, $\Delta \tilde{n}_j$, subject to the symmetrized perturbation $\hat{v}_j^\text{pert} = \tilde{\varphi}_j$. The symmetrization operation, i.e., the multiplication by $v^{1/2}$, ensures that the response can be diagonalized and also simplifies the expression of $\bar{\chi}$ in terms of $\bar{\chi}_0$. To obtain $\bar{\chi}_0$, the response is computed using the independent particle approximation, i.e., by neglecting variations to the Hartree, exchange and correlation potentials. In practice, the linear variation of the electron density may be computed using either linear response or a finite-field method~\cite{finite_ma_2019,interoperability_govoni_2021}.

In this work, we focus on the case where the charge density response is evaluated within linear response using DFPT~\cite{dfpt_baroni_1987,dfpt_baroni_2001}. In essence, for each perturbation $\hat{v}_j^\text{pert} = \tilde{\varphi}_j$, we compute the linear variation, $\Delta \psi_{i\sigma}^j$, of each occupied state of the unperturbed system, $\psi_{i\sigma}$, using the Sternheimer equation~\cite{sternheimer_sternheimer_1954}
\begin{equation}
\label{eq:sternheimer}
(h_\text{KS}^\sigma - \varepsilon_{i\sigma}) P_\text{c}^\sigma \Delta \psi_{i\sigma}^j = -P_\text{c}^\sigma v_j^\text{pert} \psi_{i\sigma} \,.
\end{equation}
Here, $P_\text{c}^\sigma$ is the projector onto the conduction (i.e., unoccupied) KS states. The completeness relation, i.e., $P_\text{c}^\sigma=1-P_\text{v}^\sigma$, where $P_\text{v}^\sigma$ is the projector onto the valence states, ensures that equation~\ref{eq:sternheimer} can be solved without explicit summations over empty states~\cite{dfpt_baroni_2001}. A preconditioned conjugate gradient method is used to solve equation~\ref{eq:sternheimer}. In practice, we note that equation~\ref{eq:sternheimer} lends itself to a nearly embarrassingly parallel implementation because it can be solved independently for each perturbation, spin channel, and orbital. Finally, the linear variation of the density caused by the $j$th perturbation is obtained as
\begin{equation}
\label{eq:dfpt}
\Delta n_j (\boldsymbol{r}) = \sum_\sigma \sum_{i=1}^{N_\text{occ}^\sigma} \Big [ \psi_{i\sigma}^* (\boldsymbol{r}) \Delta \psi_{i\sigma}^j (\boldsymbol{r}) + \Delta \psi_{i\sigma}^{j*} (\boldsymbol{r}) \psi_{i\sigma} (\boldsymbol{r}) \Big ] \,.
\end{equation}
In this way, the calculation of the response function scales as $N_\text{occ}^2 \times N_\text{PDEP} \times N_\text{PW}$, which is more favorable than conventional implementations based on the Adler-Wiser formula~\cite{dielectric_adler_1962,dielectric_wiser_1963} that scales as $N_\text{occ} \times N_\text{empty} \times N_\text{PW}^2$, where $N_\text{occ}$ ($N_\text{empty}$) is the number of occupied (empty) states. Here we use $N_\text{PW}$ as the number of plane-waves needed to represent the wave function (previously defined as $N_\psi$). The PDEP basis set allows us to achieve a low-rank decomposition of density-density response matrices, reducing the size of the matrices from $N_\chi^2$ to $N_\text{PDEP}^2$ (with $N_\text{PDEP} \ll N_\chi$). In practice, $N_\text{PDEP}$ is the only parameter of the method, and ad hoc energy cutoffs to truncate, for instance, the response function or the number of empty states are completely sidestepped. A recent verification study~\cite{gw100_govoni_2018} showed that $N_\text{PDEP}$ is just a few times the number of electrons and $N_\text{PDEP} \ll N_\text{PW}$.

WEST solves the nonlinear equation~\ref{eq:e_qp} using a root finding algorithm, e.g., the secant method, and implements the full-frequency integration in equation~\ref{eq:sigmacIR}. $G_0$ and $W_0$ are evaluated at multiple frequencies using Lanczos chains~\cite{lanczos_umari_2010,pdep_nguyen_2012}. For instance, using equation~\ref{eq:Wp} in equation~\ref{eq:Ifac} we obtain that
\begin{equation}
I_{i\sigma}(\omega) = I^{LR}_{i\sigma}(\omega)+I^{SR}_{i\sigma}(\omega) \,,
\end{equation}
where the long-range (LR) contribution and short-range (SR) contributions are:
\begin{equation}
I^{LR}_{i\sigma}(\omega) = \int_{-\infty}^{+\infty} \frac{d\omega^\prime}{2\pi}\frac{\Xi(i\omega^\prime) }{\varepsilon_{i\sigma}-\omega-i\omega^\prime} \,,
\end{equation}
\begin{equation}
I^{SR}_{i\sigma}(\omega) = \frac{1}{\Omega} \sum_{nm}^{N_\text{PDEP}} \int_{-\infty}^{+\infty} \frac{d\omega^\prime}{2\pi}\Lambda_{nm} (i\omega^\prime) \bra{\psi_{i\sigma} \tilde{\varphi}_n}(h_\text{KS}^\sigma-\omega-i\omega^\prime)^{-1}\ket{\psi_{j\sigma} \tilde{\varphi}_m} \,.
\label{eq:ISR}
\end{equation}

The shifted-inverted problem in the RHS of equation~\ref{eq:ISR} is computed introducing the Lanczos vectors:
\begin{equation}
\ket{q^{l+1}_{m i\sigma}} = \frac{1}{\beta^{l+1}_{m i\sigma}}\left[(h_\text{KS}^\sigma-\alpha^l_{m i\sigma})\ket{q^{l}_{m i\sigma}} - \beta^{l}_{m i \sigma}\ket{q^{l-1}_{m i\sigma}}\right] \,\,\, \forall l \ge 1 \,,
\end{equation}
where
\begin{equation}
\ket{q^{0}_{m i\sigma}} = 0 \,, \, \,
\ket{q^{1}_{m i\sigma}} = \ket{\psi_{i\sigma} \tilde{\varphi}_m} \,,
\end{equation}
\begin{equation}
\alpha^{l}_{m i\sigma} = \bra{q^{l}_{m i\sigma}} h_\text{KS}^\sigma \ket{q^{l}_{m i\sigma}} \,,
\end{equation}
\begin{equation}
\beta^{l+1}_{m i\sigma} = \left\lVert (h_\text{KS}^\sigma -\alpha^l_{m i \sigma}) \ket{q^{l}_{m i\sigma}} - \beta^{l}_{m i \sigma} \ket{q^{l-1}_{m i\sigma}} \right\rVert \,.
\end{equation}
By defining $d^l_{m i \sigma}$ and $U^{ll^\prime}_{m i \sigma}$ as the eigenvalues and the eigenvectors of the tri-diagonal matrix that has $\alpha^l_{m i \sigma}$ along the diagonal and $\beta^l_{m i \sigma}$ along the sub- and super-diagonal, we hence arrive at the following equation:
\begin{equation}
I^{SR}_{i\sigma}(\omega) = \frac{1}{\Omega} \sum_{nm}^{N_\text{PDEP}} \int_{-\infty}^{+\infty} \frac{d\omega^\prime}{2\pi}\Lambda_{nm} (i\omega^\prime) \sum_{ll^\prime}^{N_\text{Lanczos}} \braket{q^{1}_{n i \sigma} | q^{l}_{m i \sigma}} U ^{ll^\prime}_{m i \sigma} \frac{1}{d^{l^\prime}_{m i \sigma}-\omega -i\omega^\prime} U^{1l^\prime}_{m i \sigma} \,.
\label{eq:lanczos}
\end{equation}
In equation~\ref{eq:lanczos} we see that the dependence of $I^{SR}$ on the frequency $\omega$ is known analytically, i.e., the $U$ and $d$ coefficients and the integral in brakets do not depend on the frequency. This enables us to easily evaluate frequency-dependent quantities, which facilitates the solution of equation~\ref{eq:e_qp} without linearization, i.e., beyond the on-the-mass-shell approximation, and without using plasmon-pole models, i.e., with full-frequency. Moreover, the Lanczos vectors are obtained using a recursive algorithm that orthogonalizes newly generated vectors against previous ones. Each chain of vectors can be computed individually for each perturbation, spin channel, and orbital, resulting in a nearly embarrassingly parallel implementation.

\section{GPU Acceleration of the WEST Code}
\label{sec:gpu}

In this work, we present the porting to GPUs of the WEST code, focusing on the complete full-frequency G$_0$W$_0$ workflow shown in figure~\ref{fig:workflow}, including the construction of the PDEP basis set (with the standalone \texttt{wstat} application), the computation of $G_0$ and $W_0$ using the Lanczos algorithm, the integration of the self-energy using contour deformation, and the final solution of the QP energy levels (with the standalone \texttt{wfreq} application). For future reference, the CPU-only and GPU-accelerated versions of the WEST code are hereafter referred to as WEST-CPU and WEST-GPU, respectively.

\begin{figure}
\includegraphics[width=0.99\textwidth]{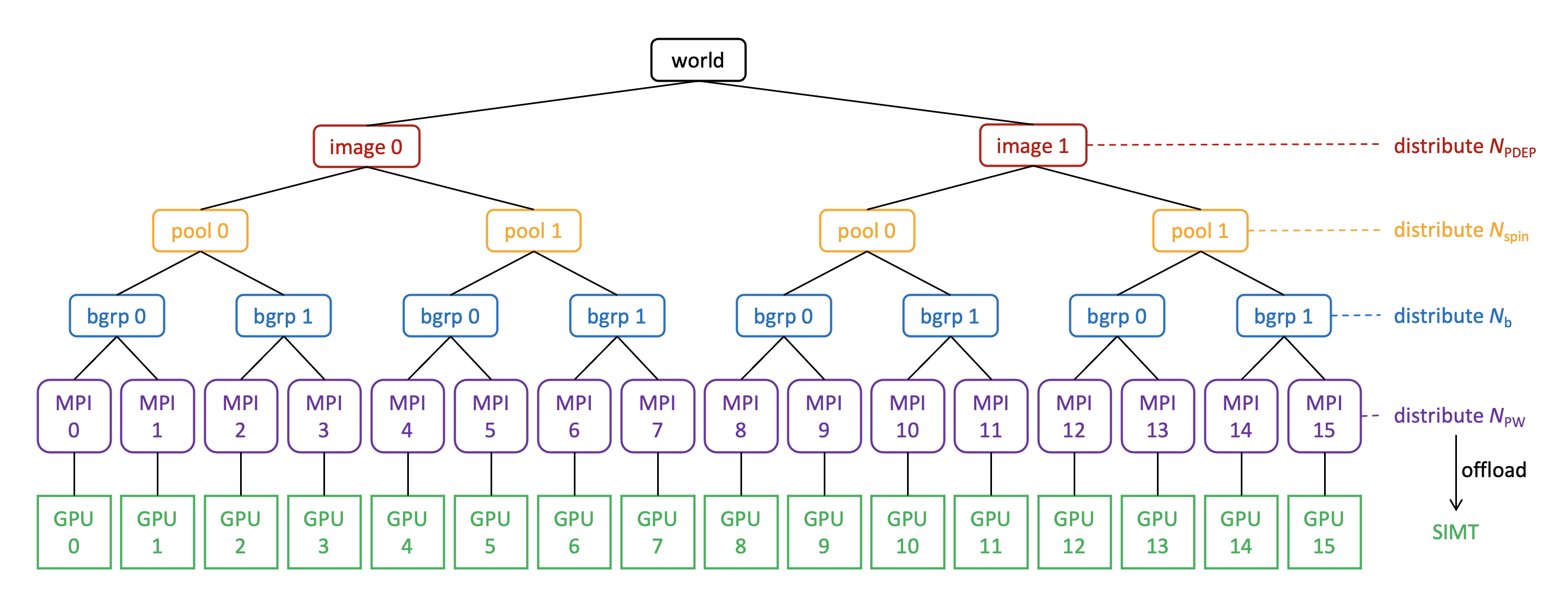}
\caption{Multilevel parallelization of the WEST code exemplified for the case of 16 total MPI processes. The processes are divided into two images. Each image is divided into two pools, each of which is further divided into two band groups. Within each band group there are two MPI processes, each of which is capable of offloading computations to a GPU device using the single instruction, multiple threads (SIMT) protocol.}
\label{fig:parallel}
\end{figure}

To meet the challenge posed by heterogeneous computing, we increased the number of parallelization levels implemented within the code, so that we can harness the embarrassingly parallel parts of the algorithms implemented in WEST as well as the data parallelism offered by GPU devices. For instance, the PDEP algorithm may be solved using $N_\text{proc}$ MPI processes and GPUs by implementing a multilevel parallelization strategy, as summarized in figure~\ref{fig:parallel}. The first level of parallelization, already introduced in reference~\citenum{west_govoni_2015}, divides $N_\text{proc}$ processes into $N_\text{image}$ subgroups, called images. Perturbations are distributed across images using a block-cyclic data distribution scheme. Each image contains a copy of the DFT data structures, such as the KS single-particle wave functions, and is responsible for computing the density response only for the perturbations owned by the image. The second and third parallelization levels, newly introduced in this work, further split the processes within an image into $N_\text{pool}$ and $N_\text{bgrp}$ subgroups, called pools and band groups, respectively. Each pool and band group stores and manipulates only a subset of the wave functions by distributing the spin polarization (for spin-polarized systems only) and band indices, respectively. The remaining $N_\text{proc}/(N_\text{image}N_\text{pool}N_\text{bgrp})$ processes within a band group distribute the plane-wave coefficients of wave functions and densities, forming the fourth level of parallelization. Finally, each MPI process is capable of offloading instructions to one GPU, which offers single instruction, multiple thread (SIMT) parallelization by leveraging the processing cores within the GPU device.

This flexible parallelization scheme helps optimize MPI communications as well as fit hardware constraints (e.g., number of GPUs within one node). Global MPI communications involving all MPI processes are avoided, except for the broadcast of the input parameters (a few scalars) from one MPI process to the others.

For offloading data-parallel regions to the GPU, we specifically focused this initial study to target NVIDIA devices. Wherever applicable, mathematical operations are performed on the GPU by calling optimized CUDA libraries, such as cuFFT for fast Fourier transforms and cuBLAS for matrix-matrix multiplications and other basic linear algebra operations. If a compute loop cannot be organized to use an existing library function, the loop is offloaded using CUDA Fortran kernel directives, which automatically generate CUDA kernels from regions of annotated CPU code~\cite{cuda_fortran}. In order to avoid the performance degradation caused by frequent data transfer operations between the CPUs and the GPUs, WEST-GPU copies the necessary data from the CPU to the GPU at the very beginning of a calculation. The data is copied back to the CPU only when absolutely necessary, e.g., for input/output (I/O) operations.

Work is underway to extend the current implementation to other GPU devices as more software and hardware become available. We anticipate that the multilevel parallelization strategy introduced so far will grant flexibility of distributing the computational workload also on GPU devices other than NVIDIA ones. However, a discussion of the performance portability and how it may be achieved by translating CUDA Fortran into OpenMP directives~\cite{openmp} goes beyond the scope of this manuscript.

In the next subsections, we elaborate on specific optimization strategies introduced in WEST-GPU on top of the multilevel parallelization. In section~\ref{sec:fft}, we point out key factors that maximize the performance of GPU-accelerated fast Fourier transforms (FFTs). In section~\ref{sec:diagonalization}, we benchmark various eigensolver libraries, identifying the most efficient solver for diagonalizing large matrices on multiple GPUs. In section~\ref{sec:async}, we demonstrate that the overhead of MPI communications can be diminished by overlapping communications with computations.

\subsection{Fast Fourier Transforms}
\label{sec:fft}

The performance of FFTs is of crucial importance to the overall efficiency of any plane-wave based electronic structure code. FFTs are extensively used in WEST to express quantities such as wave functions, densities and perturbations in either the reciprocal or the direct space. Most importantly, the application of the KS Hamiltonian to a trial wave function, a key step for the calculation of $G_0$ and $W_0$ without explicit summations over empty states, is implemented using the dual-space technique, i.e., the kinetic operator and the local potential are applied in the reciprocal or direct space, respectively. The dual-space technique takes advantage of the convolution theorem and the fact that FFTs scale as $\mathcal{O}(N\log(N))$. It follows that at least two FFTs (one forward and one backward) are required at every application of the KS Hamiltonian, and their performance greatly impacts the overall time-to-solution of both \texttt{wstat} and \texttt{wfreq} (see figure~\ref{fig:workflow}). FFTs are also invoked in other parts of the code, for example, to obtain the electron density.

WEST uses the FFTXlib library to implement parallel 3D FFTs. This library retains only the Fourier components that correspond to a chosen kinetic energy cutoff and is part of the Quantum ESPRESSO distribution~\cite{qe_giannozzi_2020}. FFTXlib may perform a 3D FFT using one MPI process or using several MPI processes by decomposing the 3D grid into slabs or pencils. The slab decomposition partitions the 3D grid into slabs, completing a 3D FFT by a set of 2D FFTs, an all-to-all communication, and a set of one-dimensional (1D) FFTs. The pencil decomposition partitions the 3D grid into pencils, completing a 3D FFT as a set of 1D FFTs, an all-to-all communication, another set of 1D FFTs, another all-to-all communication, and a final set of 1D FFTs~\cite{heffte_ayala_2020}. When multiple MPI processes are used, the Fourier components are distributed among the processes avoiding data duplication. 3D, 2D or 1D FFTs on a single MPI process are performed using vendor-optimized libraries. FFTXlib supports a variety of backends, currently including FFTW3, Intel MKL, and IBM ESSL for CPUs and cuFFT for NVIDIA GPUs.

\begin{figure}
\includegraphics[width=0.99\textwidth]{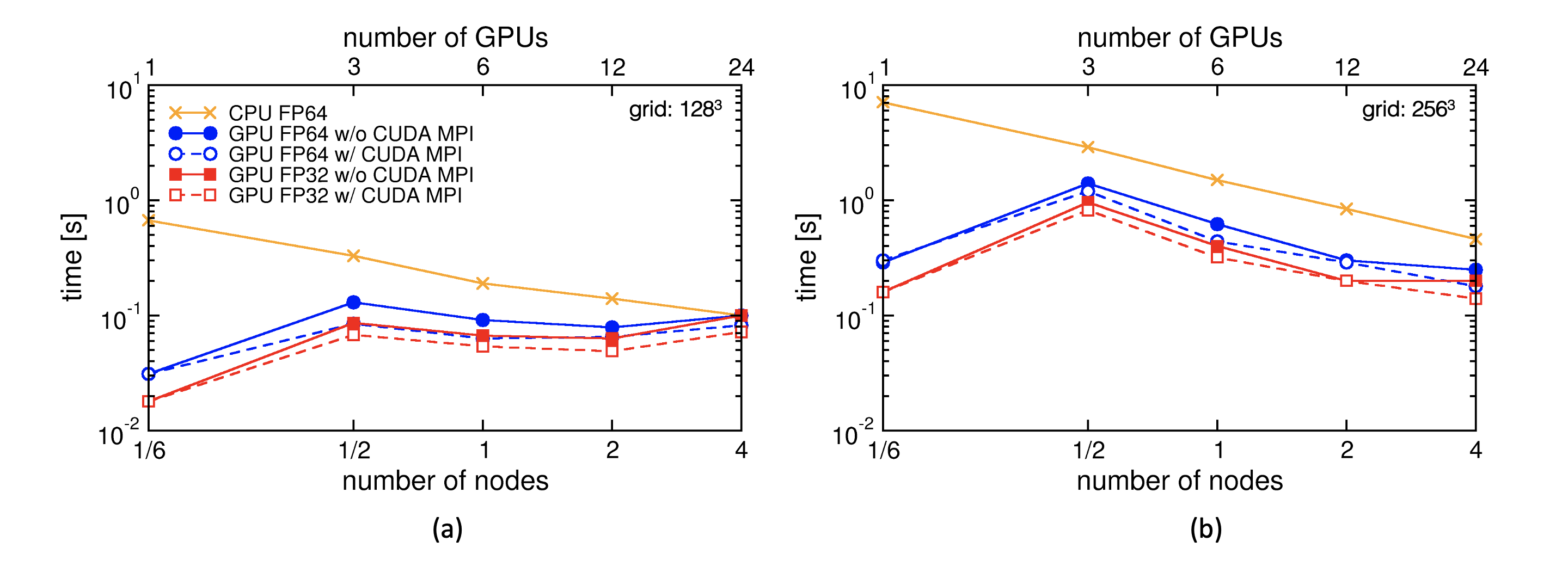}
\caption{Time required to carry out a complex-to-complex FFT operation on a 128$^3$ grid (a) or on a 256$^3$ grid (b), using the FFTXlib library on Summit (each node has six GPUs, see also the specification listed in section~\ref{sec:performance}). Blue circles and red squares identify the execution on GPUs using the cuFFT backend with double-precision (FP64) or single-precision (FP32), respectively. Timing results do not include the time needed to initially copy the data to the GPU and the time to copy the final result back to the CPU. Orange crosses identify the execution on CPUs using the ESSL backend with FP64. Filled symbols identify the use of CUDA-aware MPI and GPUDirect. Open symbols identify the use of conventional, non-CUDA-aware MPI. The slab decomposition scheme was employed for parallel FFTs.}
\label{fig:fft}
\end{figure}

We benchmarked the performance of the FFTXlib library on the Summit supercomputer located at Oak Ridge National Laboratory. Each node of Summit has two IBM POWER9 CPUs (21 cores each) and six NVIDIA V100 GPUs (see also the specification of Summit listed in section~\ref{sec:performance}). We used FFTXlib (version 6.8) with IBM ESSL (version 6.3.0) for the CPU backend and cuFFT (version 10.2.1.245) for the GPU backend. In figure~\ref{fig:fft}, we report the time needed to perform one double-precision (FP64) or single-precision (FP32) complex-to-complex (C2C) FFT for a $128^3$ or $256^3$ cubic grid using up to four nodes of Summit. Each data point in figure~\ref{fig:fft} represents the average value of 100 tests. When the FFT is GPU-accelerated, we used one MPI process per GPU. In WEST-GPU, data is pre-allocated on the GPU so that GPU-enabled FFT operations act on data that resides on the GPU. The majority of the data is therefore initialized on the GPU at the beginning of the calculation with a CPU-to-GPU copy, then the data undergoes multiple FFT operations, and finally the data is copied back to the CPU. Hence, because data transfer operations are decoupled from FFTs, figure~\ref{fig:fft} does not include the time needed to initially copy the data to the GPU and the time to copy the final result back to the CPU. We can see that one GPU (corresponding to 1/6 of a Summit node in figure~\ref{fig:fft}) outperforms one CPU (21 cores, 1/2 of a node) by more than an order of magnitude. However, while the time-to-solution on CPUs decreases linearly by increasing the number of nodes, one GPU is still faster than any other number of GPUs. Using three GPUs, for instance, slows down the calculation by nearly an order of magnitude compared to using only one GPU. Further increasing the number of GPUs leads to a moderate speedup, especially for the $256^3$ grid. However, even 24 GPUs (four nodes) cannot outperform one GPU; this is consistent with previously published benchmarks on the same machine~\cite{heffte_ayala_2020} and clearly reveals the high cost of communications (relative to the computation) involved in parallel distributed FFTs. As expected, by using FP32 instead of FP64, FFT operations achieve a $\sim$2x speedup on one GPU (red lines in figure~\ref{fig:fft}). The speedup gradually decreases as the GPU count increases, which may be attributed to the increasingly high MPI overhead relative to the small amount of computation being performed. From figure~\ref{fig:fft}, we conclude that FFTs should be carried out on as few GPUs as possible, ideally only one GPU, for maximum efficiency.

We note that the slab decomposition is used throughout this manuscript. The implementation of 3D FFTs with pencil decomposition is $\sim$30\% slower than the slab decomposition for the system sizes considered in figure~\ref{fig:fft}. In cases where one is interested in using more MPI processes than the number of slabs, the pencil decomposition will potentially become advantageous as it enables more parallelism than the slab decomposition does. This case is, however, unlikely to be relevant as figure~\ref{fig:fft} suggests that the least number of GPUs shall be used to perform FFTs due to the overhead of all-to-all communications compared to the cost of computation.

It is worth mentioning that for denser cubic grids, the overhead associated with all-to-all communications may become negligible compared to the amount of computation that needs to be performed on the GPU. A nearly ideal strong scaling of GPU-accelerated FFTs using the heFFTe library has been reported for a grid size of $1024^3$~\cite{heffte_ayala_2020}. However, the large-scale applications reported within this manuscript are performed using grids with up to 216 points per axis. Tests using the heFFTe library for smaller grids, such as $128^3$ and $256^3$, reveal performance characteristics that are similar to those of FFTXlib.

The multilevel parallelization introduced in figure~\ref{fig:parallel} is key to run WEST with as many GPUs as possible while performing FFTs using the least amount of GPUs. Specifically, the FFTs in WEST-GPU are carried out using $N_\text{proc}/(N_\text{image}N_\text{pool}N_\text{bgrp})$ MPI processes instead of all the $N_\text{proc}$ processes. In practice, $N_\text{image}$, $N_\text{pool}$, and $N_\text{bgrp}$ are chosen to restrict FFTs to the smallest number of GPUs, so that the total GPU memory is sufficiently large to accommodate the simulation data.

In the case where FFT operations involve more than one GPU, figure~\ref{fig:fft} shows that a performance gain can be obtained by taking advantage of CUDA-aware MPI and GPUDirect. Without CUDA-aware MPI, data residing on the GPU must be explicitly copied to the host CPU in order to participate in an MPI communication. If the data is needed by the GPU after the MPI communication, the data must be explicitly copied back. With CUDA-aware MPI, data on the GPU can be directly passed to MPI functions. However, depending on the hardware and software settings, the data may still be communicated through the CPU. The GPUDirect technology enhances data movement between NVIDIA GPUs. Specifically, for GPUs directly connected with each other through NVLink~\cite{nvlink_foley_2017}, the data transfer takes advantage of the high bandwidth of NVLink without going through the CPU; similarly, for internode communications, GPU data can be directly put onto the node interconnect. In figure~\ref{fig:fft}, the dashed lines correspond to FFTs employing CUDA-aware MPI and GPUDirect. For the grid sizes considered in figure~\ref{fig:fft}, switching on CUDA-aware MPI and GPUDirect results in a performance improvement ranging from 20\% to 50\%.

\subsection{Solution of Large Eigenvalue Problems}
\label{sec:diagonalization}

As introduced in section~\ref{sec:west}, WEST relies on the Davidson algorithm~\cite{davidson_davidson_1975} to iteratively diagonalize the irreducible density-density response function. In each iteration, a Hermitian matrix needs to be explicitly diagonalized. The dimension of the matrix is proportional to $N_\text{PDEP}$, and by default, matrices up to $(4 N_\text{PDEP})^2$ are diagonalized. WEST-CPU is capable of treating systems containing a few thousand electrons, leading to eigenvalue problems as large as $10000^2$. Solving such eigenvalue problems by serial or multithreaded solvers from the LAPACK library accounts for a negligible fraction of the total computational cost of WEST-CPU. For WEST-GPU, given that the most compute-intensive operations have been moved to GPUs, the eigenvalue problem stands out as roadblock that limits the performance of the code for large systems with 10000s of electrons. For instance, the largest GW calculation reported in section~\ref{sec:large} has $N_\text{PDEP} = 10368$, requesting the diagonalization of matrices up to $(4 \times 10368)^2 = 41472^2$. Solving such large eigenvalue problems on CPUs takes a significant amount of time (see table~\ref{tab:eigen}).

To circumvent this bottleneck, we compared the performance on CPUs and GPUs of four eigensolvers on Summit, namely, the multithreaded LAPACK implementation in the IBM ESSL library (version 6.3.0), the MPI-parallel and memory-distributed eigensolver in the ScaLAPACK library (version 2.1.0), the GPU-accelerated eigensolver in the cuSOLVER library (version 10.6.0.245), and the MPI-parallel, memory-distributed, GPU-accelerated eigensolver in the ELPA library (version 2020.11.001)~\cite{elpa_yu_2021}. ESSL and ScaLAPACK used one node (one MPI process, 42 OpenMP threads) and eight nodes (42 MPI processes per node), respectively. cuSOLVER and ELPA used one NVIDIA V100 GPU and eight nodes (six MPI processes per node, totaling 48 CPU cores and 48 GPUs), respectively. Table~\ref{tab:eigen} shows the performance of each eigensolver for matrix sizes ranging from $10000^2$ to $40000^2$. Using only one GPU, cuSOLVER exhibits a significant speedup over both ESSL and ScaLAPACK for matrix size up to $20000^2$. For larger matrix sizes, however, the available GPU memory (16 GB for the V100 GPU on Summit) can no longer accommodate the matrix and the workspace required by cuSOLVER. In such scenarios, the memory-distributed, GPU-accelerated ELPA eigensolver provides the fastest time-to-solution at a relatively low memory cost per GPU. On the basis of these benchmarks, WEST-GPU uses cuSOLVER to diagonalize matrices smaller than $8000^2$ and switches to GPU-accelerated ELPA for larger matrices.

Other multi-GPU eigensolvers, not considered in this work, include, for instance, SLATE~\cite{slate} and cuSOLVER-MG~\cite{cusolver_mg}. At present, SLATE is limited to compute the eigenvalues only. The commonly used 2D block-cyclic matrix distribution is not yet supported in cuSOLVER-MG, which only supports 1D block-cyclic distribution. We plan to continue assessing the performance and compatibility of these libraries as they evolve.

\begin{table}
\centering
\caption{Time to solve real symmetric eigenproblems on Summit using eigensolvers in the ESSL (multithreaded), ScaLAPACK (MPI-parallel), cuSOLVER (GPU-accelerated), and ELPA (GPU-accelerated, MPI-parallel) libraries. Matrix size $n^2$ corresponds to a square matrix with $n$ rows and $n$ columns. ``OOM'' (out of memory) indicates failed cuSOLVER executions due to insufficient device memory.}
\footnotesize
\begin{tabular}{c c c c c}
\hline
\hline
\multirow{3}{*}{Matrix size} & \multicolumn{4}{c}{Time [s]} \\
\cline{2-5}
& ESSL & ScaLAPACK & cuSOLVER & ELPA \\
& (42 CPU cores) & (336 CPU cores) & (1 GPU) & (48 GPUs) \\
\hline
10000$^2$ & \s129.1 & \s\s48.1 & \s3.2 & \s3.3 \\
20000$^2$ &  1050.4 &  \s350.6 &  22.6 & \s8.5 \\
30000$^2$ &  3483.4 &   1012.2 &   OOM &  18.4 \\
40000$^2$ &  7811.5 &   2181.0 &   OOM &  30.8 \\
\hline
\hline
\end{tabular}
\label{tab:eigen}
\end{table}

\subsection{Overlapping Computation and Communication}
\label{sec:async}

Communication overheads are reduced using nonblocking MPI functions to overlap computation and communication. Nonblocking MPI functions immediately return control to the host even if the communication has not been completed; in this way, the host is allowed to perform other operations while the communication continues in the background. When using GPUs, MPI communications can be overlapped with GPU computations and CPU-GPU communications.

\begin{figure}
\includegraphics[width=0.99\textwidth]{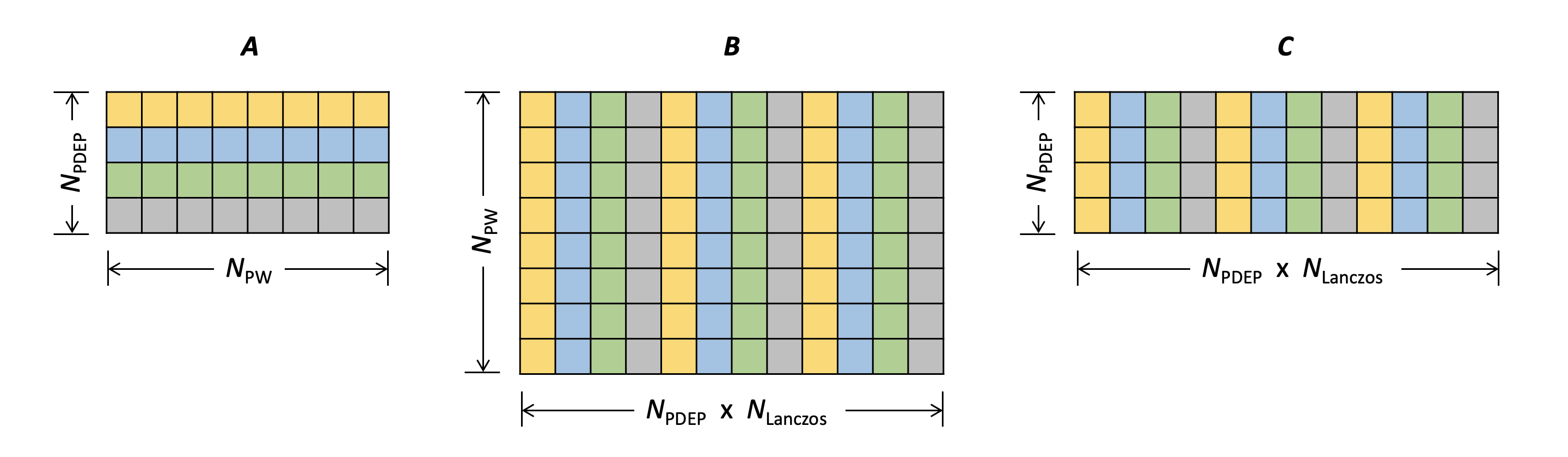}
\caption{Schematic visualization of the data distribution in the multiplication ($\boldsymbol{C} = \boldsymbol{A} \times \boldsymbol{B}$) of the two distributed matrices discussed in the text. $N_\text{PW}$ and $N_\text{PDEP}$ denote the numbers of plane-waves and PDEPs, respectively. $N_\text{Lanczos}$ denotes the length of the Lanczos chain. $N_\text{Lanczos} = 30$ typically yields converged results. Data is color-coded so that the data owned by different MPI processes have different colors.}
\label{fig:braket}
\end{figure}

Nonblocking MPI calls are extensively utilized in WEST-GPU. One example is the calculation of the braket integral in the RHS of equation~\ref{eq:lanczos}. This term may be evaluated for the $i$th orbital in the $\sigma$ spin polarization as the matrix-matrix multiplication $\boldsymbol{C} = \boldsymbol{A} \times \boldsymbol{B}$ depicted in figure~\ref{fig:braket}, where $\boldsymbol{A}_{nk}$ is the $k$th coefficient of the Fourier expansion of the product $\psi_{i\sigma}(\boldsymbol{r})\tilde{\varphi}_n(\boldsymbol{r})$ and $\boldsymbol{B}_{kml}$ is the $k$th coefficient of the Fourier expansion of $q^l_{mi\sigma}(\boldsymbol{r})$. According to figure~\ref{fig:parallel}, the indices $n$ and $m$ are distributed using the image parallelization, whereas the Fourier coefficients are distributed using the MPI processes within one band group. This distribution can lead to tall-and-skinny matrices on each process, i.e., one of the dimension is significantly greater than the other. The multiplication of tall-and-skinny matrices is memory-bound. It performs poorly on both CPUs and GPUs, which is a well-known outstanding problem. The flexible multilevel parallelization scheme reported in figure~\ref{fig:parallel} allows us to tune the shape of the local matrices, which is implemented by carefully choosing the number of images and band groups. As a result, tall-and-skinny local matrices can be avoided, pushing the matrix multiplication into the compute-bound regime and therefore achieving better performance.

The distributed matrix multiplication is completed as follows. First, each MPI process computes the product of its local portion of $\boldsymbol{A}$ and $\boldsymbol{B}$, contributing to a portion of $\boldsymbol{C}$. Second, the $i$th MPI process sends its local portion of $\boldsymbol{A}$ to process $(i-1)$ and receives another portion of $\boldsymbol{A}$ from process $(i+1)$. This communication pattern is known as a circular shift. There is no need to communicate $\boldsymbol{B}$ or $\boldsymbol{C}$. These steps are repeated until all elements of $\boldsymbol{C}$ are obtained.

\begin{figure}
\includegraphics[width=0.99\textwidth]{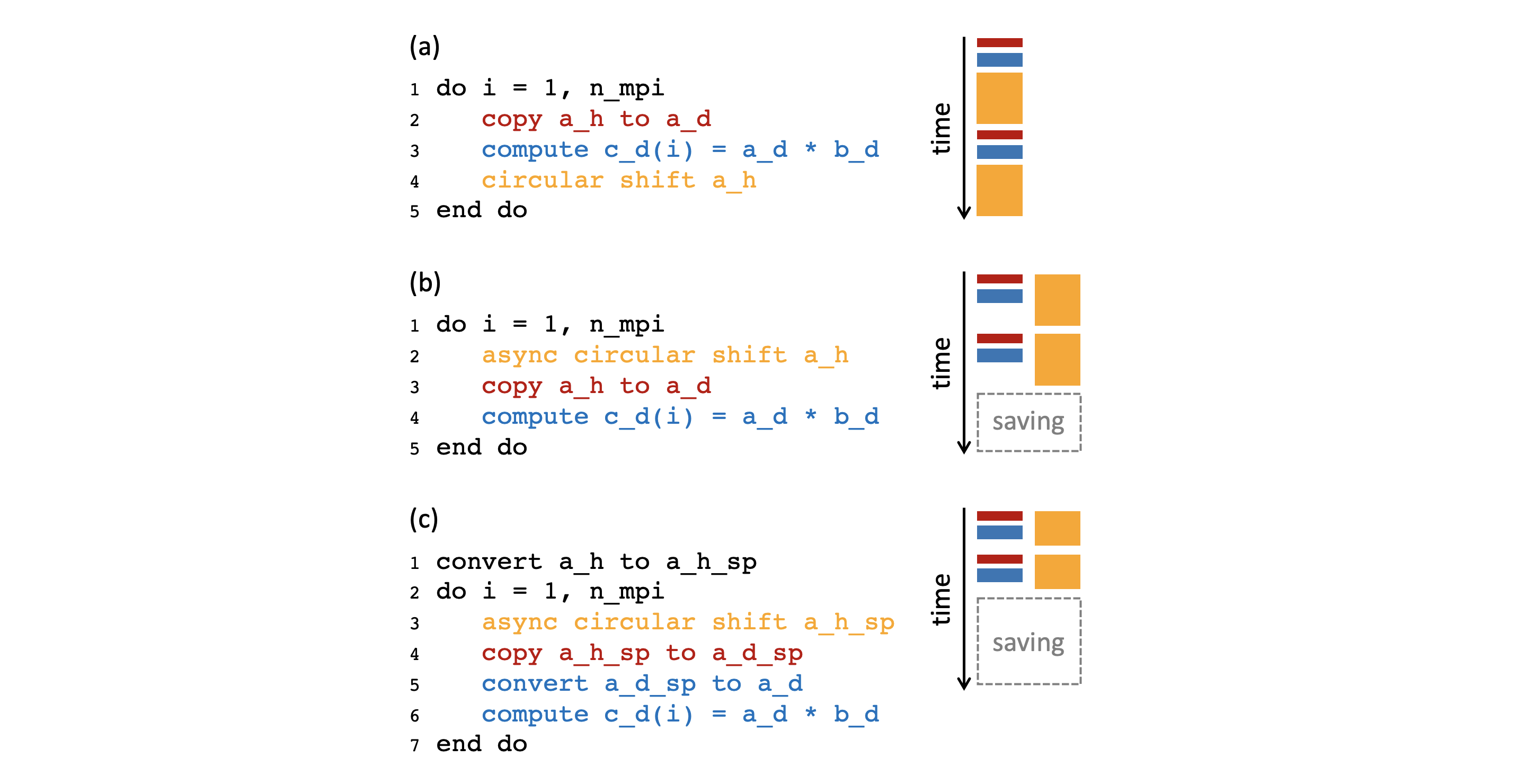}
\caption{Pseudocodes representing three alternative strategies to multiply the two distributed matrices discussed in the text. (a) No overlap between computation and communication. (b) MPI communication, CPU-GPU communication, and GPU computation are overlapped. (c) MPI communication, CPU-GPU communication, and GPU computation are overlapped, with MPI communication carried out in single-precision (FP32). Suffixes ``\_h'' and ``\_d'' indicate arrays allocated on the host (CPU) and the device (GPU), respectively. Suffix ``\_sp'' indicates a single-precision copy of a double-precision array. CPU-GPU communications, MPI communications, and GPU computation are reported in red, yellow, and blue, respectively.}
\label{fig:braket_code}
\end{figure}

The pseudocode in figure~\ref{fig:braket_code} (a) is a straightforward GPU implementation of the above procedure, which includes three sequential steps, copying $\boldsymbol{A}$ to the GPU, computing the local matrix multiplication on the GPU, and communicating $\boldsymbol{A}$ via MPI. The right part of figure~\ref{fig:braket_code} (a) shows the timeline corresponding to the pseudocode. Using nonblocking MPI functions, MPI communications can be overlapped with other operations. As shown in figure~\ref{fig:braket_code} (b), while the GPU is computing a portion of $\boldsymbol{C} = \boldsymbol{A} \times \boldsymbol{B}$, MPI communications take place asynchronously in the background to prepare the next portion of $\boldsymbol{A}$. As such, the cost of the CPU-GPU data transfer and the GPU matrix multiplication can be hidden behind the more expensive MPI communication, as the timeline in figure~\ref{fig:braket_code} indicates. In practice, this optimization leads to a speedup of 15\%-30\%.

The matrix-matrix multiplication operation in figure~\ref{fig:braket_code} can be further accelerated by performing MPI communications in single precision instead of double precision. The pseudocode of our implementation is reported in figure~\ref{fig:braket_code} (c), where $\boldsymbol{A}$ is truncated from FP64 to FP32, communicated in FP32, then cast back to FP64 and multiplied with $\boldsymbol{B}$. The precision conversions and matrix multiplications take place on the GPU, and the MPI communications are launched asynchronously to overlap with the CPU-GPU data transfers, data conversions, and matrix multiplications, as indicated by the timeline in the right part of figure~\ref{fig:braket_code} (c). The QP energies obtained using FP64 are in good agreement with the results obtained using mixed precision (FP32/FP64).

\section{Performance of WEST-GPU}
\label{sec:performance}

We report an assessment of the performance of WEST-GPU over WEST-CPU and of its strong and weak scaling using leadership HPC systems. Our benchmarks are carried out on the Summit supercomputer at Oak Ridge National Laboratory, the Perlmutter supercomputer~\bibnote{{The Perlmutter supercomputer has two phases: phase I with GPU-accelerated nodes and phase II with CPU-only nodes. Throughout this section, ``Perlmutter'' refers to the GPU-accelerated nodes in phase I.}} at the National Energy Research Scientific Computing Center, and the Theta supercomputer at Argonne National Laboratory. While the nodes of the first two supercomputers have GPUs, the nodes of the latter have only CPUs. The specifications of these computers are listed in table~\ref{tab:computers}.

\begin{table}
\centering
\caption{Specifications (per node) of the Summit, Perlmutter, and Theta supercomputers. Theoretical peak performance (TFLOP/s) is reported for double precision. The Fortran compiler and performance-critical libraries used in the benchmark calculations are also listed.}
\scriptsize
\begin{tabular}{l c c c}
\hline
\hline
& Summit & Perlmutter & Theta \\
\hline
\multirow{2}{*}{CPU} & 2 $\times$ IBM POWER9 & 1 $\times$ AMD EPYC Milan & 1 $\times$ Intel Knights Landing \\
& (2 $\times$ 21 cores) & (64 cores) & (64 cores) \\
\hline
GPU & 6 $\times$ NVIDIA V100 & 4 $\times$ NVIDIA A100 & none \\
\hline
TFLOP/s & 43.5 & 39.0 & 2.7 \\
\hline
Compiler & nvfortran 21.7 & nvfortran 21.7 & ifort 19.1.0.166 \\
\hline
& IBM Spectrum MPI 10.4 & Cray MPICH 8.1.9 & Cray MPICH 7.7.14 \\
Libraries & NVIDIA HPC SDK 21.7 & NVIDIA HPC SDK 21.7 & Intel MKL 2020 initial release \\
& CUDA 11.0.3 & CUDA 11.0.3 & \\
\hline
\hline
\end{tabular}
\label{tab:computers}
\end{table}

We conduct benchmarks of the two standalone parts of the WEST code, namely \texttt{wstat} and \texttt{wfreq}, which compute the static dielectric matrix and the full-frequency G$_0$W$_0$ self-energy, respectively (see section~\ref{sec:west}). Quantum ESPRESSO 6.8 is used for all ground-state DFT calculations. We use the SG15~\cite{oncv_schlipf_2015} optimized norm-conserving Vanderbilt (ONCV) pseudopotentials~\cite{oncv_hamann_2013} and the PBE exchange-correlation functional~\cite{pbe_perdew_1996}. As the test systems considered here are either isolated structures or large cells of periodic structures, the Brillouin zone is sampled only at the $\Gamma$-point. In \texttt{wstat} the size of the PDEP basis set is set equal to the number of electrons in the system. In \texttt{wfreq} we compute the full-frequency G$_0$W$_0$ self-energy for a variable number of states.

Benchmarks conducted on Summit and Perlmutter use CUDA-aware MPI and GPUDirect. As discussed in section~\ref{sec:fft}, these technologies facilitate the data exchange between GPUs. Timing results reported in this section correspond to the total wall clock time, including the time spent on I/O operations and CPU-GPU communications.

\subsection{Performance of WEST-GPU Relative to WEST-CPU}
\label{sec:speedup}

\begin{figure}
\includegraphics[width=0.99\textwidth]{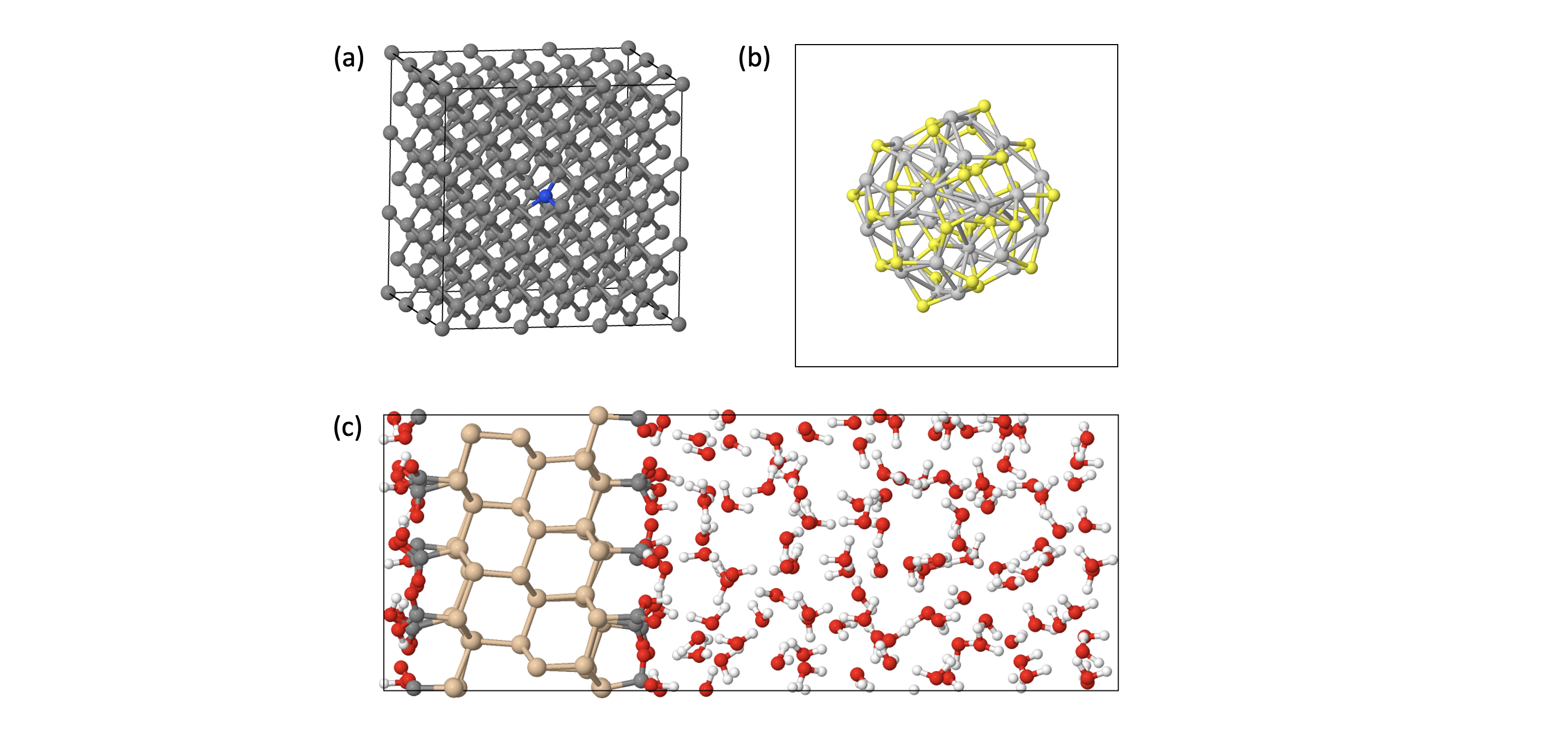}
\caption{Benchmark systems: (a) negatively-charged nitrogen-vacancy center in diamond (NV\_DIA), (b) cadmium selenide nanoparticle (CdSe\_NP), and (c) COOH-Si/H$_2$O solid/liquid interface (S/L). For the ball-and-stick atomic structures, the C, N, Cd, Se, Si, O, H atoms are colored in dark gray, blue, yellow, light gray, beige, red, and white, respectively. Details of the systems are reported in table~\ref{tab:testcases}.}
\label{fig:testcases}
\end{figure}

We compare the performance of WEST-GPU relative to the performance of WEST-CPU considering three benchmark systems: a negatively charged nitrogen-vacancy center in diamond with 215 atoms (NV\_DIA)~\cite{embedding_ma_2020a}, a Cd$_{34}$Se$_{34}$ nanoparticle (CdSe\_NP), and a COOH-Si/H$_2$O solid/liquid interface consisting of a total of 492 atoms (S/L)~\cite{west_govoni_2015,interface_pham_2014} (see figure~\ref{fig:testcases}). Details about each system are summarized in table~\ref{tab:testcases}. Because the peak performance of one node of Summit or of Perlmutter is considerably higher than the peak performance of one node of Theta (see table~\ref{tab:computers}), we benchmark WEST-GPU using 16 Summit or 16 Perlmutter nodes against the performance of WEST-CPU using 256 Theta nodes to have similar total peak performances ($\sim$696 TFLOP/s on 16 Summit nodes, $\sim$624 TFLOP/s on 16 Perlmutter nodes, $\sim$691 TFLOP/s on 256 Theta nodes).

\begin{table}
\centering
\caption{Simulation parameters for the systems depicted in figure~\ref{fig:testcases}. $N_\text{atom}$, $N_\text{electron}$, $N_\text{spin}$, and $N_\text{PW}$ denote the numbers of atoms, electrons, spin channels, and plane-waves, respectively. $E_\text{cut}$ denotes the cutoff energy used in the calculations.}
\footnotesize
\begin{tabular}{c c c c c c}
\hline
\hline
System & $N_\text{atom}$ & $N_\text{electron}$ & $N_\text{spin}$ & $E_\text{cut}$ [Ry] & $N_\text{PW}$ \\
\hline
NV\_DIA  &  215 & \s862 & 2 & 60 & \s64973 \\
CdSe\_NP & \s68 & \s884 & 1 & 50 &  382323 \\
S/L      &  492 &  1560 & 1 & 60 &  295387 \\
\hline
\hline
\end{tabular}
\label{tab:testcases}
\end{table}

\begin{figure}
\includegraphics[width=0.99\textwidth]{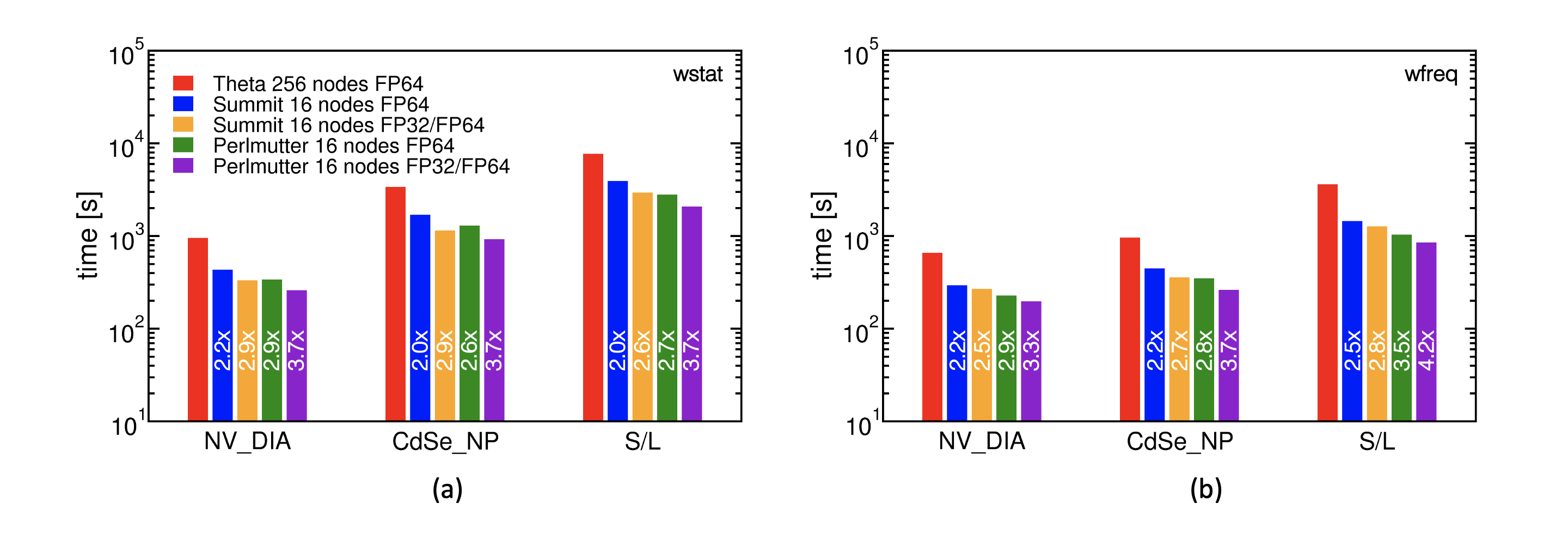}
\caption{Performance of \texttt{wstat} (a) or \texttt{wfreq} (b) using 16 nodes of Summit, 16 nodes of Perlmutter, or 256 nodes of Theta. Performance (reported within the bars) is obtained by taking the ratio of the time measured for WEST-GPU to the time measured for WEST-CPU. The \texttt{wstat} and \texttt{wfreq} codes are described in section~\ref{sec:west}. Timing results correspond to the total wall clock time, including the time spent on I/O operations and CPU-GPU communications. Forty quasiparticle energies (around the Fermi level, 20 below and 20 above) of the atomic structures in figure~\ref{fig:testcases} were calculated. Calculations on Theta used the double-precision (FP64) WEST-CPU code. Calculations on Summit and Perlmutter used the FP64 and mixed-precision (FP32/FP64) WEST-GPU code (see text).}
\label{fig:speedup}
\end{figure}

Figure~\ref{fig:speedup} (a) and (b) show the performance of the GPU accelerated \texttt{wstat} and \texttt{wfreq} parts of WEST, respectively. Using only double precision (FP64 in figure~\ref{fig:speedup}), WEST-GPU on 16 Summit nodes outperforms WEST-CPU on 256 Theta nodes by a factor of 2.0x-2.2x. This imputes an effective 32x-35x speedup for one Summit node over one Theta node, which is higher than the value of 16x, estimated by taking the ratio between the two theoretical peak performances. This may be attributed to two factors: (i) the higher node count on Theta than on Summit, which generates more internode communication, and (ii) the use of GPUs on Summit to carry out FFTs, which are notoriously memory-bound operations and therefore benefit from the higher memory bandwidth of the GPU (900 GB/s in V100 GPUs, whereas each KNL node on Theta has 16 GB fast memory with a bandwidth of 400 GB/s). By running WEST-GPU on 16 Perlmutter nodes we observe an additional 30\%-40\% speedup over 16 Summit nodes. This is caused by the fact that FFTs are carried out using one MPI process (one GPU) on Perlmutter, while on Summit we are forced to use three MPI processes (three GPUs) due to the memory limitation (40 GB in A100 GPUs, 16 GB in V100 GPUs). FFTs in the latter case incur the overheads described in section~\ref{sec:fft}. Moreover, the A100 GPU has a higher memory bandwidth (1555 GB/s) than the V100 GPU and features FP64 tensor cores that can be automatically utilized by the CUDA libraries wherever possible. On the contrary, the V100 GPU features tensor cores only for half precision, which are not utilized by the current version of WEST. Similar conclusions are drawn analyzing the FP64 performance of \texttt{wfreq}, where we computed 40 QP energies (around the Fermi level, 20 below and 20 above) for each system.

Figure~\ref{fig:speedup} also reports the performance of the mixed-precision (FP32/FP64) version of WEST-GPU. In the case of mixed precision, the code operates in FP64 except for the regions of the code with distributed matrix multiplication or FFTs, which are carried out using FP32, as discussed in section~\ref{sec:gpu}. The FP32/FP64 code outperforms the FP64 counterpart on both Summit and Perlmutter by up to 45\%, due to a nearly two-fold speedup in the corresponding FFT and MPI operations. It is important to note that the QP energies obtained using the FP32/FP64 code are in good agreement with the results obtained with FP64. The mean absolute error of the 40 QP energies computed with the FP32/FP64 code lies well below $10^{-4}$ eV for the three systems studied here, justifying the utilization of mixed precision in production calculations. For all calculations reported in sections~\ref{sec:scaling} and \ref{sec:large}, the FP32/FP64 version of WEST was employed.

Table~\ref{tab:flops_testcases} reports the performance of WEST-GPU in terms of FLOP/s, computed as the ratio of the total number of FLOPs to the total time of the simulation. FLOPs are counted by inserting counters into the source code. This approach comes with a lower overhead than using external profiling tools. Nevertheless, we measure the FLOPs and the time in two separate runs in order to obtain accurate timing results. The performance of WEST-GPU is compared against the theoretical peak performance of Summit and Perlmutter. For both \texttt{wstat} and \texttt{wfreq}, a higher fraction of the theoretical peak was reached on Perlmutter, ranging from 36.0\% to 72.9\%, than on Summit, ranging from 23.6\% to 49.7\%. GPUs are better utilized for larger systems, as the workload associated with larger systems is more likely to saturate the GPUs. We note that the ratio to peak performance shown in table~\ref{tab:flops_testcases} is used to approximately estimate how efficiently the GPUs are being utilized. In WEST-GPU, the FFTs benefit from the use of FP32, and the matrix-matrix multiplications and possibly other linear algebra operations benefit from the FP64 tensor cores on Perlmutter. These factors are not reflected in table~\ref{tab:flops_testcases}.

\begin{table}
\centering
\caption{Performance (TFLOP/s) of WEST-GPU on 16 nodes of Summit and Perlmutter for the systems described in table~\ref{tab:testcases}. The \texttt{wstat} and \texttt{wfreq} codes are described in section~\ref{sec:west}. Timing results correspond to the total wall clock time, including the time spent on I/O operations and CPU-GPU communications. Performance (Perf.) is measured as the ratio of the total number of FLOPs ($N_\text{FLOP}$) to the total time of the simulation. ``\% Peak'' denotes the ratio of the measured performance to the theoretical peak performance; the latter is calculated as $N_\text{node} \times 43.5$ TFLOP/s and $N_\text{node} \times 39.0$ TFLOP/s for the Summit and Perlmutter computers, respectively.}
\tiny
\begin{tabular}{c c c| c c c| c c c}
\hline
\hline
\multirow{2}{*}{Code} & \multirow{2}{*}{System} & $N_\text{FLOP}$ & \multicolumn{3}{c|}{Summit} & \multicolumn{3}{c}{Perlmutter} \\
\cline{4-9}
& & [TFLOPs] & Time [s] & Perf. [TFLOP/s] & \% Peak & Time [s] & Perf. [TFLOP/s] & \% Peak \\
\hline
               & NV\_DIA  & $9.21 \times 10^4$ & \s332.0 & 277.5 & 39.9 & \s260.1 & 354.3 & 56.8 \\
\texttt{wstat} & CdSe\_NP & $3.96 \times 10^5$ &  1147.0 & 345.6 & 49.7 & \s919.2 & 431.2 & 69.1 \\
               & S/L      & $9.42 \times 10^5$ &  2937.1 & 320.6 & 47.4 &  2070.9 & 454.8 & 72.9 \\
\hline
               & NV\_DIA  & $4.41 \times 10^4$ & \s267.8 & 164.8 & 23.6 & \s196.4 & 224.7 & 36.0 \\
\texttt{wfreq} & CdSe\_NP & $1.18 \times 10^5$ & \s354.7 & 333.5 & 47.9 & \s262.7 & 450.1 & 70.1 \\
               & S/L      & $3.63 \times 10^5$ &  1269.1 & 286.0 & 41.1 & \s851.8 & 426.1 & 68.2 \\
\hline
\hline
\end{tabular}
\label{tab:flops_testcases}
\end{table}

\subsection{Strong and Weak Scaling of WEST-GPU}
\label{sec:scaling}

\begin{figure}
\includegraphics[width=0.99\textwidth]{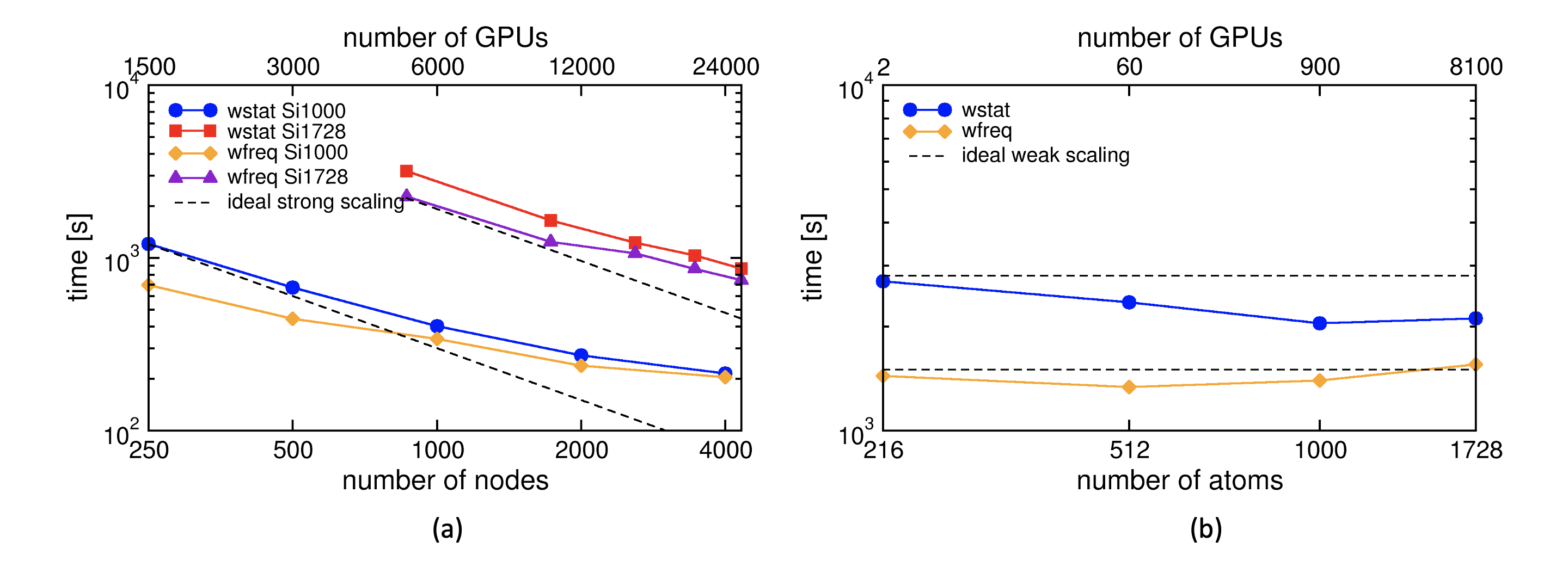}
\caption{Strong (a) and weak (b) scaling of WEST-GPU. The \texttt{wstat} and \texttt{wfreq} codes are described in section~\ref{sec:west}. (a) Blue circles (yellow diamonds) represent the strong scaling of \texttt{wstat} (\texttt{wfreq}) for the 1000-atom silicon supercell reported in table~\ref{tab:silicon}; red squares (violet triangles) represent the strong scaling of \texttt{wstat} (\texttt{wfreq}) for a 1728-atom silicon supercell reported in table~\ref{tab:silicon}. (b) Blue circles (yellow diamonds) represent the weak scaling of \texttt{wstat} (\texttt{wfreq}). Black dashed lines indicate the slope of ideal scaling. Eighty quasiparticle energies (around the Fermi level, 40 below and 40 above) were calculated for each system. Timing results correspond to the total wall clock time, including the time spent on I/O operations and CPU-GPU communications.}
\label{fig:scaling}
\end{figure}

We report the strong and weak scaling of WEST-GPU as benchmarked on the Summit supercomputer with a series of silicon supercell models with up to 1728 atoms, as described in table~\ref{tab:silicon}. The strong scaling of WEST-GPU is presented in figure~\ref{fig:scaling} (a) for two Si supercells containing 1000 or 1728 atoms. The weak scaling is presented in figure~\ref{fig:scaling} (b) for four Si supercells containing 216, 512, 1000, and 1728 atoms. Eighty QP energies (around the Fermi level, 40 below and 40 above) were calculated for each system. Strong and weak scaling close to the ideal one (dashed lines) is observed for both \texttt{wstat} and \texttt{wfreq}. The 1728-atom system exhibits a strong scaling closer to the ideal one than that of the 1000-atom system. This stems from the higher computation-to-communication ratio of the larger system, and it demonstrates the applicability of WEST-GPU to large-scale simulations.

\begin{table}
\centering
\caption{Simulation parameters of the silicon supercells used as benchmarks. Supercells are obtained by considering replicas of the eight-atom conventional unit cell. $N_\text{atom}$, $N_\text{electron}$, $N_\text{spin}$, and $N_\text{PW}$ denote the numbers of atoms, electrons, spin channels, and plane-waves, respectively. $E_\text{cut}$ denotes the cutoff energy used in the calculations.}
\footnotesize
\begin{tabular}{c c c c c c}
\hline
\hline
Supercell & $N_\text{atom}$ & $N_\text{electron}$ & $N_\text{spin}$ & $E_\text{cut}$ [Ry] & $N_\text{PW}$ \\
\hline
$3 \times 3 \times 3$ & \s216 & \s864 & 1 & 16 & \s31463 \\
$4 \times 4 \times 4$ & \s512 &  2048 & 1 & 16 & \s74773 \\
$5 \times 5 \times 5$ &  1000 &  4000 & 1 & 16 &  145837 \\
$6 \times 6 \times 6$ &  1728 &  6912 & 1 & 16 &  251991 \\
\hline
\hline
\end{tabular}
\label{tab:silicon}
\end{table}

In table~\ref{tab:flops_silicon} we report an estimate of the performance of WEST-GPU by measuring the total number of floating-point operations recorded for running \texttt{wstat} and \texttt{wfreq} and dividing it by the total time, including the time to carry out I/O operations. For the 1000-atom silicon supercell, \texttt{wstat} reaches 47.3\% and 16.6\% of the theoretical peak performance on 250 and 4000 Summit nodes, respectively. Internode MPI communications are responsible for the drop in the performance at large number of nodes. When we increase the size of the system to comprise 1728 silicon atoms, \texttt{wstat} reaches 42.5\% of the peak on 864 nodes and sustains 31.2\% of the peak even on 4320 nodes (25920 V100 GPUs), amounting to a mixed-precision performance of 58.80 PFLOP/s. The performance of \texttt{wfreq} is slightly inferior to that of \texttt{wstat} due to the larger amount of internode MPI communications in \texttt{wfreq}. Nevertheless, for the 1728-atom silicon supercell, \texttt{wfreq} achieves a mixed-precision performance of 35.88 PFLOP/s on 4320 nodes, corresponding to 19.1\% of the peak. In all cases, we observe that the full applications (including I/O operations) scale to the entire Summit machine. We note again that the ratio to peak performance is discussed for a qualitative understanding of how the GPUs are utilized by WEST-GPU. It does not take into consideration that WEST-GPU carries out FFTs and MPI communications in single precision.

\begin{table}
\centering
\caption{Performance (PFLOP/s) of WEST-GPU on Summit for the 1000 and 1728 silicon atoms supercell models described in table~\ref{tab:silicon}. The \texttt{wstat} and \texttt{wfreq} codes are described in section~\ref{sec:west}. Timing results correspond to the total wall clock time, including the time spent on I/O operations and CPU-GPU communications. Performance (Perf.) is measured as the ratio of the total number of FLOPs ($N_\text{FLOP}$) to the total time of the simulation. $N_\text{node}$ denotes the number of Summit nodes used in the calculations (each node has six GPUs, see table~\ref{tab:computers}). ``\% Peak'' denotes the ratio of the measured performance to the theoretical peak performance; the latter is calculated as $N_\text{node} \times 43.5$ TFLOP/s.}
\footnotesize
\begin{tabular}{c c c c c c c c}
\hline
\hline
Code & $N_\text{atom}$ & $N_\text{FLOP}$ [PFLOPs] & $N_\text{node}$ & Time [s] & Perf. [PFLOP/s] & \% Peak \\
\hline
\multirow{7}{*}{\texttt{wstat}} & \multirow{2}{*}{1000} & \multirow{2}{*}{$6.20 \times 10^3$} & \s250 &  1204.6 & \s5.15 &  47.3 \\
                                &                       &                                     &  4000 & \s214.1 &  28.95 &  16.6 \\
\cline{2-7}
                                & \multirow{5}{*}{1728} & \multirow{5}{*}{$5.09 \times 10^4$} & \s864 &  3182.7 &  16.01 &  42.5 \\
                                &                       &                                     &  1728 &  1648.4 &  30.89 &  41.1 \\
                                &                       &                                     &  2592 &  1226.5 &  41.52 &  36.8 \\
                                &                       &                                     &  3456 &  1030.8 &  49.40 &  32.9 \\
                                &                       &                                     &  4320 & \s866.0 &  58.80 &  31.2 \\
\hline
\multirow{7}{*}{\texttt{wfreq}} & \multirow{2}{*}{1000} & \multirow{2}{*}{$2.95 \times 10^3$} & \s250 & \s695.7 & \s4.24 &  39.0 \\
                                &                       &                                     &  4000 & \s203.5 &  14.50 & \s8.3 \\
\cline{2-7}
                                & \multirow{5}{*}{1728} & \multirow{5}{*}{$2.67 \times 10^4$} & \s864 &  2259.0 &  11.82 &  31.4 \\
                                &                       &                                     &  1728 &  1239.1 &  21.55 &  28.7 \\
                                &                       &                                     &  2592 &  1062.0 &  25.14 &  22.3 \\
                                &                       &                                     &  3456 & \s864.7 &  30.88 &  20.5 \\
                                &                       &                                     &  4320 & \s744.1 &  35.88 &  19.1 \\
\hline
\hline
\end{tabular}
\label{tab:flops_silicon}
\end{table}

For the system with 1000 silicon atoms, the FLOP count ($N_\text{FLOP}$) required to compute the quasiparticle energy for $N_\text{QP}$ bands using \texttt{wfreq} is $N_\text{FLOP} = (2553 + 8.17 \times N_\text{QP})$ PFLOPs. The prefactor (2553 PFLOPs) identifies the FLOPs required to compute the dielectric screening at all frequencies and without empty states using the PDEP basis set, while the multiplicative factor (8.17 PFLOPs) is attributed to the cost of computing the full-frequency G$_0$W$_0$ self-energy for one band. The FLOP count indicates that it becomes cost-effective to compute the self-energy for many states; this is convenient, for instance, for the simulation of photoelectron spectra over an extended region of energies~\cite{westapp_gaiduk_2016}.

Finally, we note that the FLOP count in table~\ref{tab:flops_silicon} indicates that a few EFLOPs are necessary in order to compute the full-frequency G$_0$W$_0$ electronic structure of both benchmark systems. At the measured sustained 30-60 PFLOP/s throughput, the calculations can be carried out within tens of minutes. We also note that the current results are obtained with an implementation that, in addition to avoiding approximating the screened Coulomb interaction with generalized plasmon-pole models, sidesteps altogether the need to compute many empty states using DFT and the need to introduce a stringent energy cutoff in reciprocal space to represent dielectric matrices.

\section{Large-Scale Full-Frequency G$_0$W$_0$ Calculations}
\label{sec:large}

\begin{figure}
\includegraphics[width=0.99\textwidth]{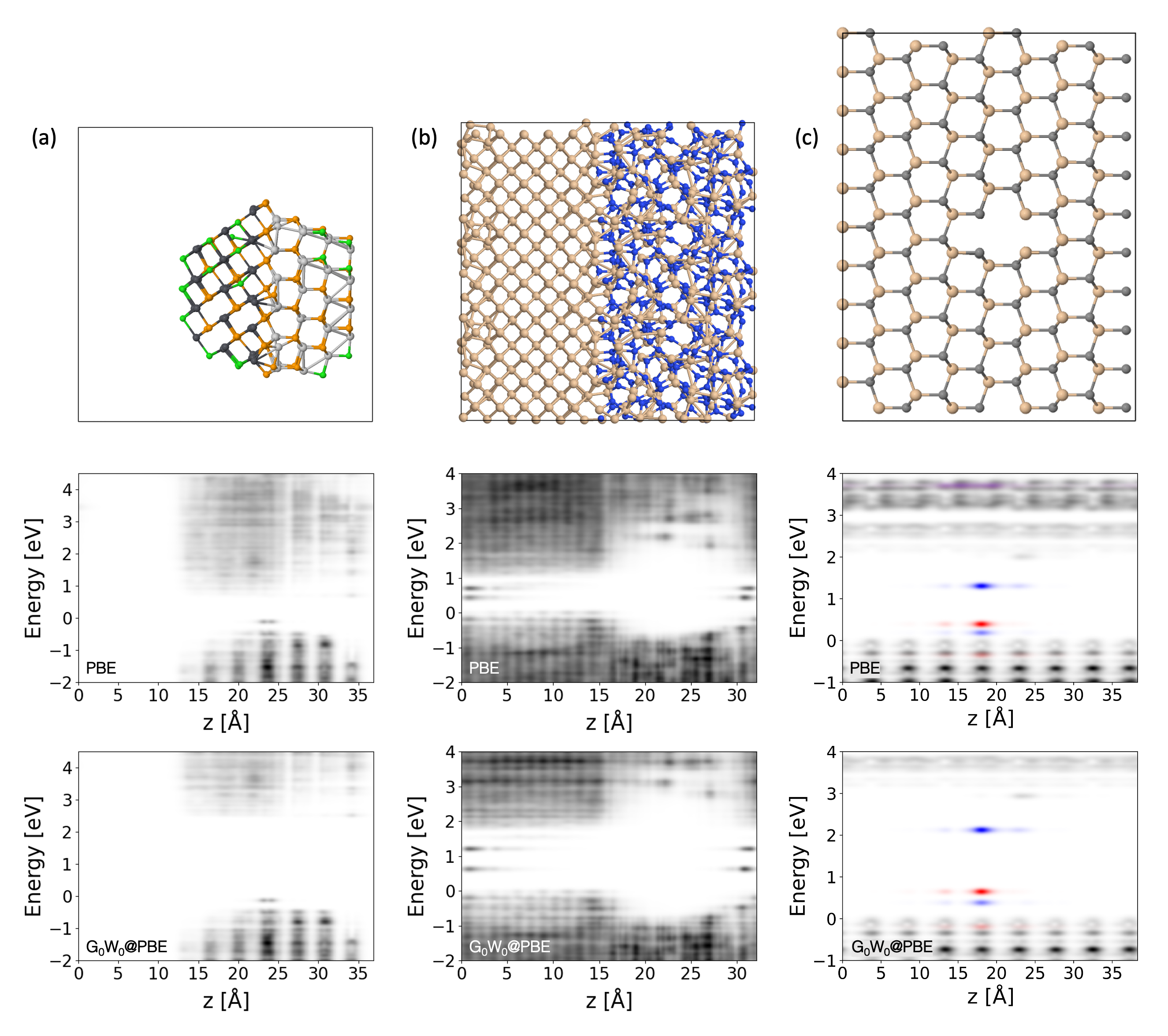}
\caption{Large-scale full-frequency G$_0$W$_0$ calculations considered in this work: (a) Janus-like heterostructure formed by a chlorine-terminated nanoparticle made of cadmium sulfide and lead sulfide (CdS/PbS), (b) interface of silicon and silicon nitride (Si/Si$_3$N$_4$), and (c) neutral hh divacancy in 4H silicon carbide (VV\_SiC). Top panels report a side view of the simulation cells. For the ball-and-stick atomic structures the Cl, Cd, S, Pb, Si, N, C atoms are colored in green, black, orange, light gray, beige, blue, and dark gray, respectively. Bottom and middle panels report the local density of states (LDOS, see text) obtained using G$_0$W$_0@$PBE or KS-DFT energies in equation~\ref{eq:ldos}, respectively. LDOS is plotted using a color scale ranging from white to black; white areas indicate energy gaps. For VV\_SiC, the defect states in the up (down) spin channel are shown in red (blue). Details of the systems are reported in table~\ref{tab:applications}.}
\label{fig:applications}
\end{figure}

Finally, we demonstrate the capability of WEST-GPU for computing the full-frequency G$_0$W$_0$ electronic structure of large-scale systems. The structures shown in figure~\ref{fig:applications} are representative examples of large heterogeneous systems of interest for energy sustainability and quantum information science research. The structure in figure~\ref{fig:applications} (a) is a Janus nanoparticle (CdS/PbS) consisting of 301 atoms and 2816 electrons. In this system, investigated for its applicability to photovoltaics~\cite{janus_kroupa_2018}, we compute with G$_0$W$_0$ the band offsets between the cadmium sulfide (CdS) and the lead sulfide (PbS) hemispheres of the heterostructured nanoparticle. The structure in figure~\ref{fig:applications} (b) is an interface model of silicon and silicon nitride (Si/Si$_3$N$_4$), which was used to model high dielectric constant materials for electronics~\cite{interface_pham_2013}. Also for this system, which has 2376 atoms and 10368 electrons, we compute the band offsets between the two materials using G$_0$W$_0$. The structure in figure~\ref{fig:applications} (c) is a neutral hh divacancy in a $10 \times 10 \times 2$ supercell of 4H silicon carbide (VV\_SiC)~\cite{westapp_jin_2021}. This is a representative system for solid-state quantum information technologies where G$_0$W$_0$ is used to identify the energy of deep defect states. The system has 1598 atoms and 6392 electrons and, unlike the previous two systems, requires an explicit treatment of spin polarization. The details of the structures in figure~\ref{fig:applications} are summarized in table~\ref{tab:applications}.

\begin{table}
\centering
\caption{Simulation parameters for the systems depicted in figure~\ref{fig:applications}. $N_\text{atom}$, $N_\text{electron}$, $N_\text{spin}$, and $N_\text{PW}$ denote the numbers of atoms, electrons, spin channels, and plane-waves, respectively. $E_\text{cut}$ denotes the cutoff energy used in the calculations.}
\footnotesize
\begin{tabular}{c c c c c c c}
\hline
\hline
System & $N_\text{atom}$ & $N_\text{electron}$ & $N_\text{spin}$ & $E_\text{cut}$ [Ry] & $N_\text{PW}$ \\
\hline
CdS/PbS        & \s301 & \s2816 & 1 & 30 & 948557 \\
Si/Si$_3$N$_4$ &  2376 &  10368 & 1 & 30 & 638633 \\
VV\_SiC        &  1598 & \s6392 & 2 & 30 & 314653 \\
\hline
\hline
\end{tabular}
\label{tab:applications}
\end{table}

For all considered systems we computed the local density of states (LDOS), defined as:
\begin{equation}
\label{eq:ldos}
\text{LDOS}(z,E) = \sum_{i\sigma} \int \frac{\text{d}x}{L_x} \int \frac{\text{d}y}{L_y} \vert \psi_{i\sigma} (x,y,z)\vert ^2 \delta(E - \varepsilon_{i\sigma}) \,,
\end{equation}
where $\psi_{i\sigma}$ and $\varepsilon_{i\sigma}$ are the wave functions and their G$_0$W$_0$ or PBE energies; $L_x$ and $L_y$ are the lengths of the x and y axes of the simulation box, respectively, whereas $z$ corresponds to the z axis of the simulation box; and $\delta$ is the Dirac delta function (modeled by a Gaussian function with a width of 0.05 eV). The middle panel of figure~\ref{fig:applications} reports the LDOS computed using PBE wave functions and energy levels. The bottom panel of figure~\ref{fig:applications} reports the LDOS computed using PBE wave functions and full-frequency G$_0$W$_0$ energy levels. To compute the LDOS, the QP energies of 480, 2000, and 1200 single-particle states were computed for the CdS/PbS, Si/Si$_3$N$_4$, and VV\_SiC structures, respectively. As expected, the LDOS at the G$_0$W$_0$@PBE level exhibits a larger energy gap compared to the PBE result for all structures. For the Janus-like nanoparticle and the Si/Si$_3$N$_4$ interface, the LDOS allows us to track the density of states as a function of the coordinate $z$ that is perpendicular to the interface. For the system in figure~\ref{fig:applications} (c), the energy gap of 4H-SiC obtained at the G$_0$W$_0$@PBE level, 3.17 eV, is in close agreement with the experimental value of 3.2 eV~\cite{sic_levinshtein_2001}. At the G$_0$W$_0$ (PBE) level of theory the energy difference between the $e$ and the $a_1$ defect states in the minority spin channel is $e-a_1=1.73 (1.12)$ eV. We obtain an exciton binding energy and an ionic relaxation energy of 0.45 and 0.10 eV, respectively, from a DFT calculation~\bibnote{The exciton binding energy and ionic relaxation energy are estimated from the results of three DFT calculations using the DDH functional. The occupations of the KS orbitals are constrained in these calculations. The first calculation uses the structure and occupations of the ${}^3A_2$ ground state. We obtain the total energy of the ${}^3A_2$ state, $E_\text{tot}$, and the difference between the $e$ and the $a_1$ defect states in the minority spin channel, $e-a_1$. The second calculation uses the structure of the ${}^3A_2$ state and the occupations of the ${}^3E$ excited state, resulting in a total energy $E_\text{tot}^\prime$. The total energy difference, $E_\text{tot}^\prime-E_\text{tot}$, corresponds to the vertical neutral excitation energy from the ${}^3A_2$ state to the ${}^3E$ state. The exciton binding energy $E_\text{bind}$ is estimated as $(e-a_1)-(E_\text{tot}^\prime-E_\text{tot})$. The third calculation uses the structure and occupations of the ${}^3E$ state. We obtain the total energy of the ${}^3E$ state, $E_\text{tot}^{\prime \prime}$. The total energy difference, $E_\text{tot}^{\prime \prime}-E_\text{tot}$, corresponds to the ionic relaxation energy $E_\text{relax}$.} of the hh divacancy in an $8 \times 8 \times 2$ supercell of 4H-SiC using the dielectric dependent hybrid functional (DDH)~\cite{ddh_skone_2016}. Subtracting these energies from $e-a_1$ computed at the G$_0$W$_0$ level of theory, we obtain 1.18 eV, which is close to the measured zero-phonon line (ZPL) of 1.095 eV~\cite{sic_falk_2013}.

The calculations reported in this section were carried out on the Summit supercomputer using $\sim$10000 NVIDIA V100 GPUs (six GPUs per node). The measured number of FLOPs, time-to-solution, and performance in terms of FLOP/s are shown in table~\ref{tab:flops_applications}. The \texttt{wstat} (\texttt{wfreq}) code achieves up to 35.8\% (23.2\%) of the theoretical peak performance. Due to the size of the memory available in V100 GPUs, we had to distribute FFT operations within each band group on 12 GPUs. This configuration does not yield optimal performance for FFTs (one GPU per band group), as discussed in section~\ref{sec:fft}. We anticipate to see improved performance on GPUs that have more device memory than the V100 GPUs.

\begin{table}
\centering
\caption{Performance (PFLOP/s) of WEST-GPU on Summit for the systems described in table~\ref{tab:applications}. The \texttt{wstat} and \texttt{wfreq} codes are described in section~\ref{sec:west}. Timing results correspond to the total wall clock time, including the time spent on I/O operations and CPU-GPU communications. Performance (Perf.) is measured as the ratio of the total number of FLOPs ($N_\text{FLOP}$) to the total time of the simulation. $N_\text{node}$ denotes the number of Summit nodes used in the calculations (each node has six GPUs, see table~\ref{tab:computers}). ``\% Peak'' denotes the ratio of the measured performance to the theoretical peak performance; the latter is calculated as $N_\text{node} \times 43.5$ TFLOP/s.}
\footnotesize
\begin{tabular}{c c c c c c c c}
\hline
\hline
Code & System & $N_\text{FLOP}$ [PFLOPs] & $N_\text{node}$ & Time [s] & Perf. [PFLOP/s] & \% Peak \\
\hline
\multirow{3}{*}{\texttt{wstat}} & CdS/PbS        & $1.43 \times 10^4$ & 1408 & \s\s731.9 & 19.54 & 31.9 \\
                                & Si/Si$_3$N$_4$ & $4.04 \times 10^5$ & 1728 &   18356.9 & 21.95 & 29.2 \\
                                & VV\_SiC        & $8.81 \times 10^4$ & 1600 &  \s3536.4 & 24.91 & 35.8 \\
\hline
\multirow{3}{*}{\texttt{wfreq}} & CdS/PbS        & $7.99 \times 10^3$ & 1408 & \s\s562.0 & 14.22 & 23.2 \\
                                & Si/Si$_3$N$_4$ & $2.54 \times 10^5$ & 1728 &   16309.3 & 15.58 & 20.7 \\
                                & VV\_SiC        & $6.39 \times 10^4$ & 1600 &  \s3333.0 & 19.17 & 20.9 \\
\hline
\hline
\end{tabular}
\label{tab:flops_applications}
\end{table}

\section{Conclusions}
\label{sec:conclusions}

We reported the use of GPUs to carry out large-scale full-frequency G$_0$W$_0$ calculations using WEST, a code for many-body perturbation theory calculations based on the plane-wave and pseudopotential method. Compared to other conventional implementations of G$_0$W$_0$, the algorithms implemented in WEST do not require the calculation of many empty states nor the definition of a stringent energy cutoff for response functions. In this work, we introduced a multilevel parallelization strategy in WEST that is devised to distribute the computational workload and reduce the overhead cost associated with MPI communications. We discussed a number of optimizations that improve the performance of the code on GPU-equipped supercomputers, including the use of mature high-performance libraries, and strategies to minimize data transfer operations between CPUs and GPUs. In addition, the utilization of mixed precision led to a considerable performance improvement over the solely double-precision version without sacrificing accuracy.

The GPU-accelerated version of WEST realizes substantial speedup over its CPU-only counterpart and displays excellent strong and weak scaling, as benchmarked on the Summit supercomputer using up to 25920 NVIDIA V100 GPUs. The code reaches a mixed-precision (FP32/FP64) performance of 58.8 PFLOP/s. The same code runs seamlessly on the Perlmutter supercomputer equipped with NVIDIA A100 GPUs, which are one generation newer than the V100 GPUs on Summit. The newly developed GPU code has the capability of advancing the simulation of electronic excitations in large heterogeneous materials, as demonstrated by carrying out full-frequency G$_0$W$_0$ calculations of a nanostructure, an interface, and a defect in a wide band gap material using supercells with up to 10368 electrons. These calculations are carried out overcoming commonly adopted approximations, e.g., the use of generalized plasmon-pole models, and are, to the best of our knowledge, the largest deterministic full-frequency G$_0$W$_0$ calculations performed to date.

We are exploring the possibility to extend mixed-precision regions to other memory-intensive or compute-intensive portions of the code, e.g., the storage of the nonlocal part of the pseudopotential and the evaluation of the exact exchange needed in hybrid density functionals~\cite{mixed_vinson_2020}. Work is in progress to optimize the performance on GPUs of the electron-phonon~\cite{phonon_mcavoy_2018,phonon_yang_2021}, the Bethe-Salpeter equation (BSE) in finite field~\cite{absorption_nguyen_2019,absorption_dong_2021}, and the quantum embedding~\cite{embedding_ma_2020a,embedding_ma_2020b,embedding_ma_2021} kernels.

Data related to this publication are organized using the Qresp software~\cite{qresp_govoni_2019} and are available online at \url{https://paperstack.uchicago.edu}.

The authors declare no competing financial interest.

\begin{acknowledgement}
The authors acknowledge support provided by the Midwest Integrated Center for Computational Materials (MICCoM), as part of the Computational Materials Sciences Program funded by the U.S. Department of Energy, Office of Science, Basic Energy Sciences, Materials Sciences, and Engineering Division through Argonne National Laboratory (ANL). This research used resources of the Oak Ridge Leadership Computing Facility (OLCF) at the Oak Ridge National Laboratory, which is supported by the Office of Science of the U.S. Department of Energy under Contract No. DE-AC05-00OR22725. This research used resources of the National Energy Research Scientific Computing Center (NERSC), a U.S. Department of Energy Office of Science User Facility operated under Contract No. DE-AC02-05CH11231. This work was additionally supported by participation in the NERSC Exascale Application Readiness Program. This research used resources of the Argonne Leadership Computing Facility (ALCF) at ANL, a U.S. Department of Energy Office of Science User Facility operated under Contract No. DE-AC02-06CH11357.

We thank Dr. Brandon Cook (NERSC), Dr. Soham Ghosh (NERSC), Prof. Giulia Galli (University of Chicago and ANL), Prof. Fran\c{c}ois Gygi (University of California, Davis), Dr. Christopher Knight (ALCF), Ryan Prout (OLCF), Dr. Marcello Puligheddu (ANL), Dr. Jonathan Skone (NERSC), and Dr. John Vinson (National Institute of Standards and Technology) for fruitful discussions. We gratefully acknowledge the organizers of a GPU hackathon event co-organized by ALCF and NVIDIA Corporation, and Dr. William Huhn (ALCF) and Dr. Kristopher Keipert (NVIDIA) for their advice during this event.
\end{acknowledgement}

\bibliography{west}

\begin{mcitethebibliography}{129}
\bibitem[Not()]{Note-1}
{The Perlmutter supercomputer has two phases: phase I with GPU-accelerated
  nodes and phase II with CPU-only nodes. Throughout this section,
  ``Perlmutter'' refers to the GPU-accelerated nodes in phase I.}\relax
\mciteBstWouldAddEndPunctfalse
\mciteSetBstMidEndSepPunct{\mcitedefaultmidpunct}
{}{\mcitedefaultseppunct}\relax
\EndOfBibitem
\bibitem[Not()]{Note-2}
The exciton binding energy and ionic relaxation energy are estimated from the
  results of three DFT calculations using the DDH functional. The occupations
  of the KS orbitals are constrained in these calculations. The first
  calculation uses the structure and occupations of the ${}^3A_2$ ground state.
  We obtain the total energy of the ${}^3A_2$ state, $E_\text{tot}$, and the
  difference between the $e$ and the $a_1$ defect states in the minority spin
  channel, $e-a_1$. The second calculation uses the structure of the ${}^3A_2$
  state and the occupations of the ${}^3E$ excited state, resulting in a total
  energy $E_\text{tot}^\prime$. The total energy difference,
  $E_\text{tot}^\prime-E_\text{tot}$, corresponds to the vertical neutral
  excitation energy from the ${}^3A_2$ state to the ${}^3E$ state. The exciton
  binding energy $E_\text{bind}$ is estimated as
  $(e-a_1)-(E_\text{tot}^\prime-E_\text{tot})$. The third calculation uses the
  structure and occupations of the ${}^3E$ state. We obtain the total energy of
  the ${}^3E$ state, $E_\text{tot}^{\prime \prime}$. The total energy
  difference, $E_\text{tot}^{\prime \prime}-E_\text{tot}$, corresponds to the
  ionic relaxation energy $E_\text{relax}$.\relax
\mciteBstWouldAddEndPunctfalse
\mciteSetBstMidEndSepPunct{\mcitedefaultmidpunct}
{}{\mcitedefaultseppunct}\relax
\EndOfBibitem
\end{mcitethebibliography}

\end{document}